\documentclass[a4paper,11pt]{article}
\usepackage{amsfonts}
\usepackage{amsmath}% AMSLaTeX宏包 用来排出更加漂亮的公式
\usepackage{amssymb}% AMSLaTeX宏包 用来排出更加漂亮的公式
\usepackage{amsthm}%定理和证明环境
\usepackage{color,bibcheck}
\usepackage{algorithm,algorithmic}
\usepackage[section]{placeins}
\allowdisplaybreaks[4]%鼓励公式自动分页
\usepackage{mathrsfs, graphicx,hyperref}
\usepackage[all]{xy}
\usepackage{amsmath}
\UseRawInputEncoding%提交arxiv时用
\begin{document}
\title{\bf Automated Peripancreatic Vessel Segmentation and Labeling Based on Iterative Trunk Growth and Weakly Supervised Mechanism}

\author{Liwen Zou\footnote{E-mail: dz20210008@smail.nju.edu.cn. Department of Mathematics, Nanjing University, Nanjing 210008, P.R. China.}, Zhenghua Cai, Liang Mao, Ziwei Nie, Yudong Qiu\footnote{(Corresponding author) E-mail: yudongqiu510@nju.edu.cn. Department of General Surgery, Nanjing Drum Tower Hospital, Nanjing 210008, P.R. China.} and Xiaoping Yang\footnote{(Corresponding author) E-mail: xpyang@nju.edu.cn. Department of Mathematics, Nanjing University, Nanjing 210008, P.R. China.}}

\maketitle

\begin{abstract}
Peripancreatic vessel segmentation and anatomical labeling play extremely important roles to assist the early diagnosis, surgery planning and prognosis for patients with pancreatic tumors. However, most current techniques cannot achieve satisfactory segmentation performance for peripancreatic veins and usually make predictions with poor integrity and connectivity. Besides, unsupervised labeling algorithms cannot deal with complex anatomical variation while fully supervised methods require a large number of voxel-wise annotations for training, which is very labor-intensive and time-consuming. To address these problems, we propose our Automated Peripancreatic vEssel Segmentation and lAbeling (APESA) framework, to not only highly improve the segmentation performance for peripancreatic veins, but also efficiently identify the peripancreatic artery branches. There are two core modules in our proposed APESA framework: iterative trunk growth module (ITGM) for vein segmentation and weakly supervised labeling mechanism (WSLM) for artery branch identification. Our proposed ITGM is composed of a series of trunk growth modules, each of which chooses the most reliable trunk of a basic vessel prediction by the largest connected constraint, and seeks for the possible growth branches by branch proposal network. Our designed iterative process guides the raw trunk to be more complete and fully connected. Our proposed WSLM consists of an unsupervised rule-based preprocessing for generating pseudo branch annotations, and an anatomical labeling network to learn the branch distribution voxel by voxel. We achieve Dice of 94.01\% for vein segmentation on our collected dataset, which boosts the accuracy by nearly 10\% compared with the state-of-the-art methods. Additionally, we also achieve Dice of 97.01\% on segmentation and competitive performance on anatomical labeling for peripancreatic arteries.


\end{abstract}

\section{Introduction}
\begin{figure}[h]
		\centering
		{
			\includegraphics[width=9cm,]{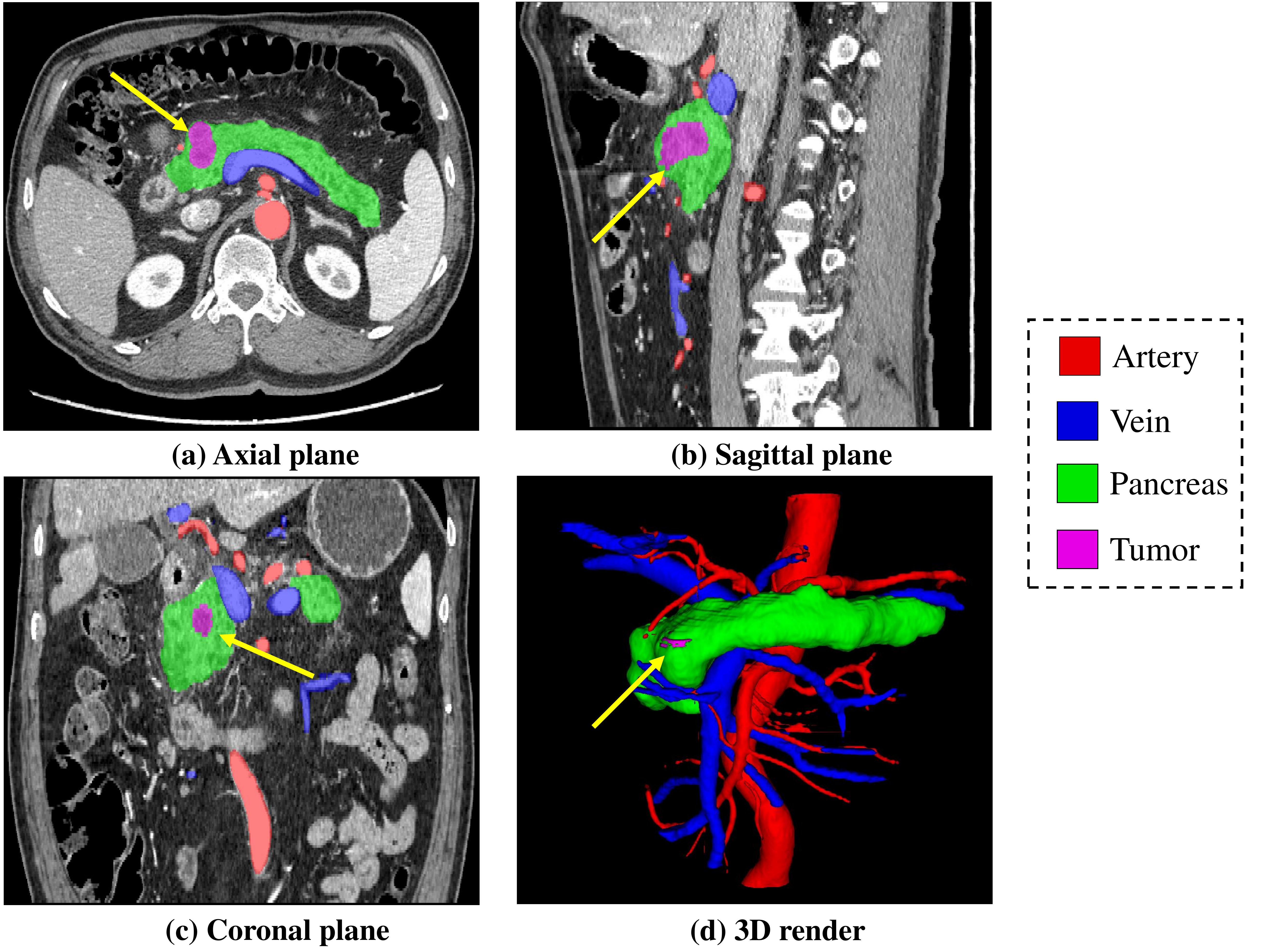}
		}
		\caption{Visualization of relationship between the pancreas, pancreatic tumor and peripancreatic vessels. The tumor areas are specially indicated by the yellow arrows.}\label{vessel-tumor}
	\end{figure}

Pancreatic cancer is the third most common cause of cancer deaths in the United States \cite{Siegel}. Pancreatic ductal adenocarcinoma (PDAC) is the most prevalent malignant pancreatic tumor, accounting for over 90\% of pancreatic malignancies \cite{Wild}, and it is the most intractable pancreatic cancer with a poor prognosis, where the operability and survival chance are strongly affected by the tumor infiltration of the surrounding vessels especially arteries. Peripancreatic vascular anatomy is complex and tumors often tend to infiltrate both arteries and veins \cite{Dima}. Figure \ref{vessel-tumor} shows the relationship between the pancreas, pancreatic tumor and peripancreatic vessels on contrast computed tomography (CT) images. Hence, automated segmentation and anatomical labeling for peripancreatic vessels are promising for the early diagnosis, surgery planning and prognosis to patients with pancreatic cancers.

\FloatBarrier

To the best of our knowledge, only one existing work concentrated on the peripancreatic vessel segmentation \cite{Dima}. This work achieved nearly perfect segmentation performance on peripancreatic arteries in multispectral CT images. However, it only provided one argument for segmenting the arteries, but the veins. And it does not focus on the anatomical labeling for the artery branches, which is extremely important for clinical issues. More and more studies demonstrate the significance of vein analysis for tumor treatment. And identifying or labeling peripancreatic artery structure is of great importance for the diagnosis and treatment of pancreatic cancers. Therefore, the goal of this work is to address the problems: peripancreatic vein segmentation and artery labeling.

\begin{figure}[h]
		\centering
		{
			\includegraphics[width=12cm,]{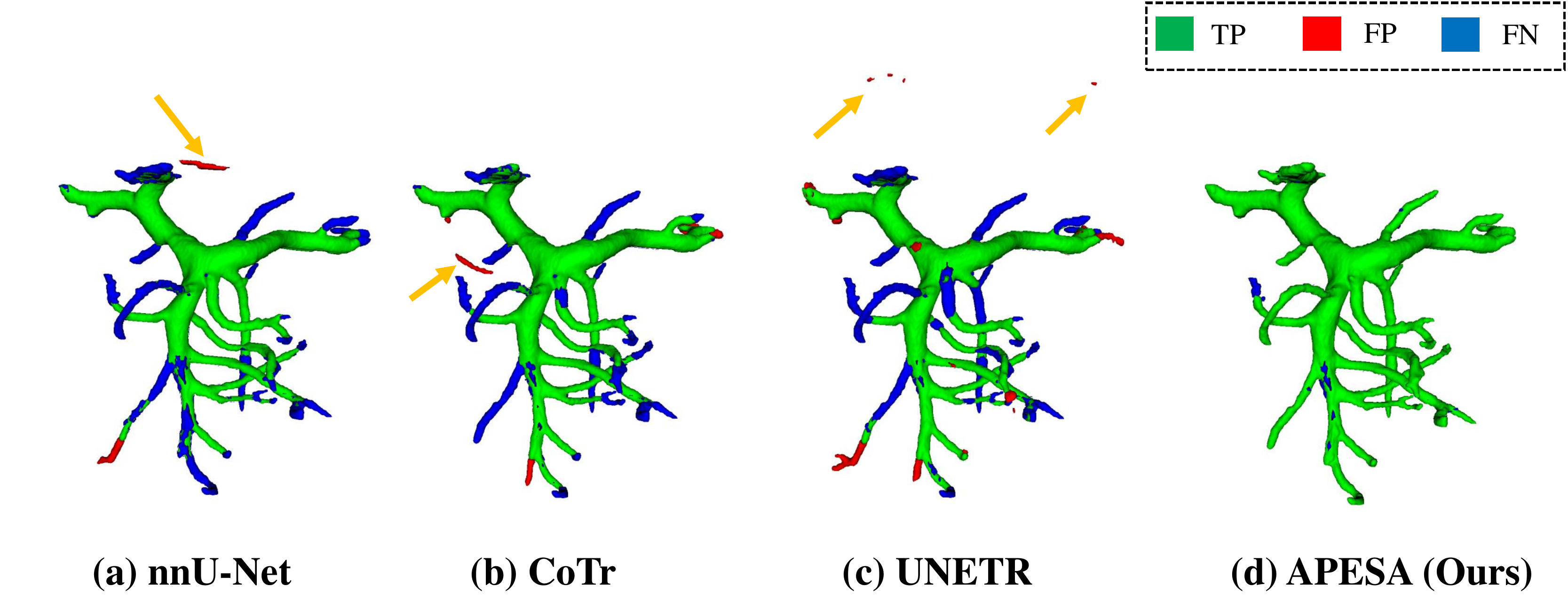}
		}
		\caption{An example to illustrate the benefits of the proposed APESA method for peripancreatic vein segmentation. (a) nnU-Net (b) CoTr (c) UNETR and (d)the proposed APESA method. The green, red and blue voxels denote the true positive (TP), false positive (FP) and false negative (FN) predictions, respectively. It can be found that the result of proposed method has fewer false negative predictions and no false negative predictions almost. Besides, our prediction is fully connected while the previous methods predict some noisy components indicated by yellow arrows. }\label{problem}
	\end{figure}

 In CT images, the peripancreatic veins on the venous phase have lower contrast than the arteries on arterial phase, which brings extra difficulty for segmentation. Besides, previous techniques show the weakness of the vein segmentation performance for the terminal integrity and connectivity. Figure \ref{problem} shows the experimental results that we compare the existing deep learning-based methods and our proposed APESA method, where we can see that existing methods can not deal with the terminal errors and disconnection well.

\FloatBarrier

\begin{figure}[h]
		\centering
		{
			\includegraphics[width=8cm,]{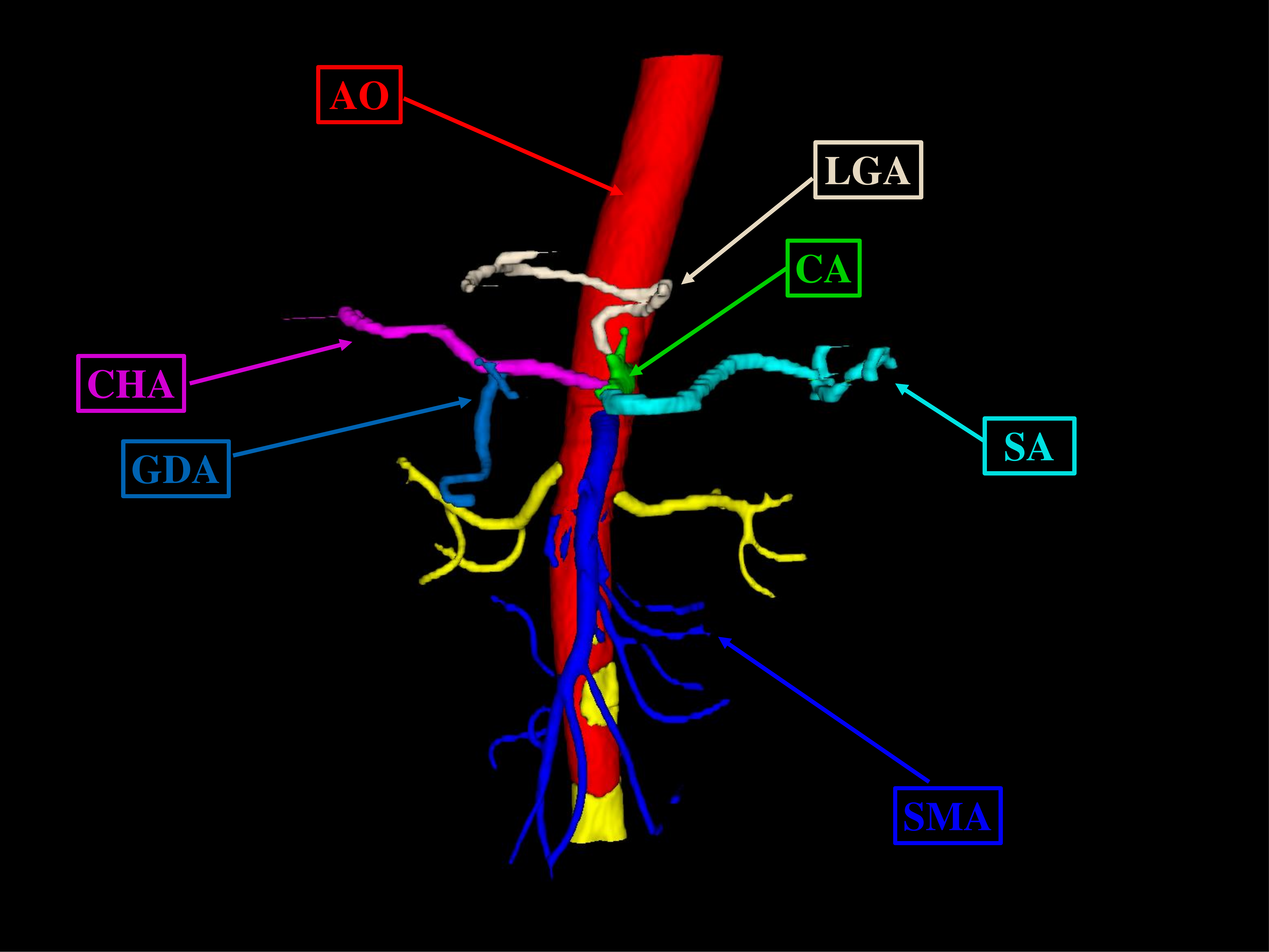}
		}
		\caption{Visualization of the anatomical labeling for peripancreatic
arteries}\label{labeling}
	\end{figure}

As for the anatomical labeling of peripancreatic arteries, there are various studies focusing on the abdominal artery anatomical labeling \cite{Oda2012,Matsuzaki2015,Liu2022}. Most of these studies adopt unsupervised rule-based algorithms or fully supervised learning-based methods. The rule-based algorithms usually cannot deal with the anatomical variation and have low efficiency, while the fully supervised learning-based methods require a large number of voxel-wise annotations, which are labor-intensive and time-consuming. In this work, we have proposed a weakly supervised anatomical labeling mechanism which aims to identify the peripancreatic artery branches with very less participation of experts. Figure \ref{labeling} shows an example of our focused peripancreatic artery branches. For clinical needs, labeling the abdominal aorta (AO), celiac artery(CA), superior mesenteric artery (SMA), splenic artery (SA), common hepatic artery (CHA), left gastric artery (LGA) and gastroduodenal artery (GDA) has the highest priority. Therefore, our goal is to identify the above branches in our experiments.

\FloatBarrier

The contributions of our work can be summarized as follows.
\begin{itemize}
\item To the best of our knowledge, this is the first work to simultaneously address the problem of peripancreatic vessel segmentation and anatomical labeling. We propose our novel \textbf{A}utomated \textbf{P}eripancreatic v\textbf{E}ssel \textbf{S}egmentation and l\textbf{A}beling (\textbf{APESA}) framework for highly improving peripancreatic vein segmentation accuracy and efficiently identifying peripancreatic artery branches.

\item We propose an iterative trunk growth module (ITGM) composed of a series of trunk growth modules. Our proposed ITGM combines the largest connected constraint and branch proposal network to guide a raw vessel trunk to be more complete and fully connected by an iterative growth process.

\item We propose a weakly supervised labeling mechanism (WSLM) to efficiently identify peripancreatic artery branches with very less participation of experts. Our proposed WSLM consists of an unsupervised rule-based preprocessing to generate pseudo artery branch labels which are judged by experts and an anatomical labeling network to learn the branch distributions voxel by voxel.

\item Experimental results show that our proposed APESA approach boosts the vein segmentation accuracy by nearly 10\% (Dice: 94.01\%) compared with the state-of-the-art (SOTA) methods, and achieves more than 20\% (clDice: 87.96\%) improvement for the topological integrity compared with the previous techniques. Besides, we also achieve the best (Dice: 97.01\%) and competitive performance on segmentation and anatomical labeling, respectively for peripancreatic arteries.

\end{itemize}

\section{Related work}

\subsection{Abdominal vessel segmentation in CT images}
In recent years, a myriad of dedicated blood vessel segmentation algorithms for different body regions and modalities have been developed \cite{Moccia2018}. Ibragimov \textit{et al.} \cite{Ibragimov} applied convolutional neural networks (CNNs) to learn the consistent appearance patterns of the portal vein (PV) and achieved Dice score of 83\% in their experiments.  Oda \textit{et al.} \cite{Oda2019} segmented abdominal arteries from an abdominal CT volume with a fully convolutional network (FCN) \cite{FCN} and they obtained 87.1\% F-measure in a public dataset BTCV \cite{BTCV}. Golla \textit{et al.} \cite{Golla} used a novel ratio-based sampling method to train 2D and 3D versions of the U-Net \cite{unet}, the V-Net \cite{vnet} and the DeepVesselNet \cite{DNet} for abdominal blood vessel segmentation, and achieved Dice score of 75.8\% and 83.8\% on veins and arteries respectively. Dima \textit{et al.} \cite{Dima} also trained a U-Net based model to perform binary segmentation of the peripancreatic arteries in multispectral CT images, where they obtained a near perfect segmentation with a Dice score of 95.05\% in their best performing model. Mahmoudi \textit{et al.} \cite{Mahmoudi} introduced a framework based on CNN for segmentation of PDAC mass and surrounding vessels in CT images by incorporating powerful classic features, and got the Dice score of 73\% and 81\% on superior mesenteric vein (SMV) and SMA, respectively. Zhu \textit{et al.} \cite{ZhuOda} added two auxiliary tasks to the FCN to extract the skeleton context of abdominal arteries and achieved Dice score of 93.2\% in the same public dataset as in \cite{Oda2019}.

\subsection{Region proposal and distraction attention}
Region proposal strategy is widely used in computer vision especially object detection tasks. Ren \textit{et al.} \cite{Ren2015} introduced a region proposal network (RPN) to efficiently generate high quality detection region proposals, and these region proposals are fed to the subsequent training classifier to be judged as true or false. The distraction concepts have been explored in many computer vision tasks, such as semantic segmentation \cite{Huang}, saliency detection \cite{Chen,Xiao} and visual tracking \cite{Zhu}. Zhao \textit{et al.} \cite{Zhao} proposed a cascaded two-stage U-Net based model to explicitly take the ambiguous region information into account. In their model, the first stage generates a global segmentation for the whole input CT volume and predicts latent distraction regions, which contain both false negative areas and false positive areas, against the segmentation ground truth. The second stage embeds the distraction region information into segmentation for volume patches to further discriminate the target regions.

Inspired by the above methods and considering the characteristics of blood vessel segmentation, we reformulate the vessel segmentation task: finding missing branches in predictions as possible based on a reliable trunk segmentation. Therefore, we develop a new branch proposal network by learning the distraction regions to provide missing branch proposals based on a raw vessel trunk prediction. Instead of feeding these proposals to a training classifier in \cite{Ren2015}, we use the most essential prior knowledge: the vessel is fully connected, to identify the rationality of the generated proposals.

\subsection{Anatomical labeling for abdominal vessels}
Anatomical labeling for abdominal vessels has been greatly developed in recent years. Oda \textit{et al.} \cite{Oda2012} presented an automated anatomical labeling method of both upper or lower abdominal arteries by designed rules and machine learning. They achieved 79.01\% and 80.41\% for recall and precision, respectively. Matsuzaki \textit{et al.} \cite{Matsuzaki2015} adopted the similar strategy combing rule-based preprocessing, machine learning techniques and rule-based postprocessing to label the abdominal arteries and a hepatic portal system. And they obtained the labeling F-measure of 85.7\% and 72.2\% on manually and automatically extracted arteries, respectively. The proposed method by Kitasaka \textit{et al.} \cite{Kitasaka2017} for vessel labeling represents a blood vessel tree as a probabilistic graphical model by conditional random fields (CRFs)\cite{CRF}, and got the F-measure of 94.4\% for abdominal arteries. Liu \textit{et al.} \cite{Liu2022} defined a hypergraph representation of the abdominal arterial system as a family tree model with a probabilistic hypergraph matching framework for automated vessel labeling. They achieved the average F-measure of 93.0\% for abdominal arteries.

\section{Methods}
\subsection{Network architecture}
\begin{figure*}[htbp]
		\centering
		{
			\includegraphics[width=15cm,]{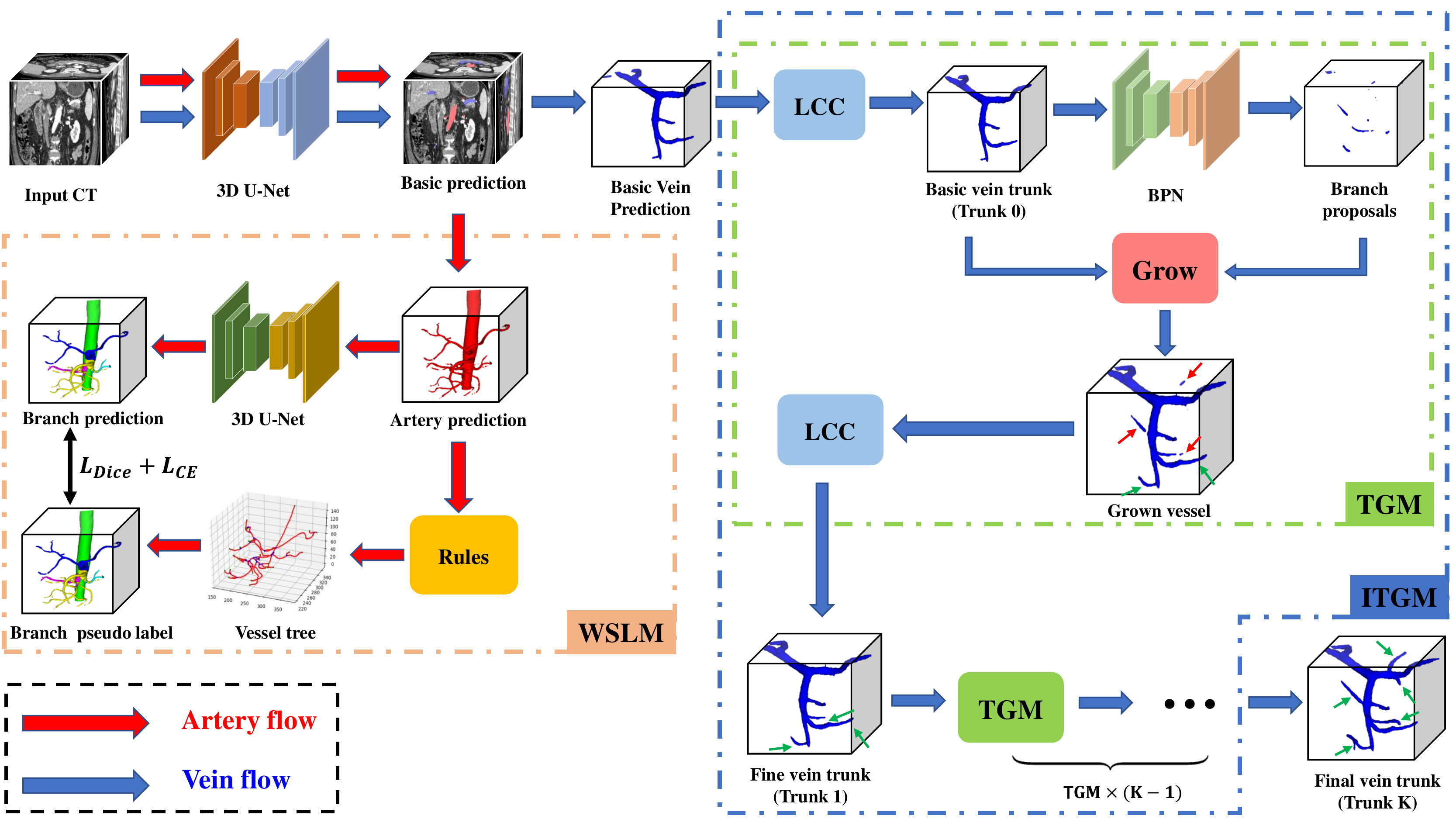}
		}
		\caption{Pipeline of our proposed \textbf{A}utomated \textbf{P}eripancreatic V\textbf{E}ssel \textbf{S}egmentation and L\textbf{A}beling (\textbf{APESA}) framework  for peripancreatic vein segmentation and artery labeling.}\label{pipeline}
	\end{figure*}

As shown in Figure \ref{pipeline}, we propose a novel framework for peripancreatic vein segmentation and artery labeling on CT images. We term it as \textbf{A}utomated \textbf{P}eripancreatic V\textbf{E}ssel \textbf{S}egmentation and L\textbf{A}beling (\textbf{APESA}) framework. There are two processing flows in our APESA method, one for arteries and another for veins. When our APESA is fed with a CT scan, a basic segmentation module composed with a U-Net generates the basic predictions for peripancreatic arteries and veins. For the subsequent vein flow, we design an iterative trunk growth module (ITGM) composed of a series of trunk growth modules (TGMs) to obtain more complete and fully connected vein predictions based on the basic segmentation results. Each TGM is composed of two largest connected constraint (LCC) operations to get reliable vessel trunks, and a branch proposal network (BPN) to find surrounding possible branches to make the trunk grow. For the subsequent artery flow, considering the importance of the peripancreatic arteries to clinical issues, and the satisfactory segmentation performance by U-Net architecture for peripancreatic arteries \cite{Dima}, we concentrate on designing the labeling algorithm.  The basic artery predictions is directly fed to the proposed weakly supervised labeling mechanism (WSLM) for branch identification. Our proposed WSLM aims to efficiently identify most important artery branches for pancreatic tumor treatment by combing rule-based pseudo label generation and learning-based anatomical labeling network.

\FloatBarrier

\subsection{Basic segmentation module}
The basic segmentation module is composed of an nnU-Net \cite{nnunet} model based on the U-Net architecture to segment the peripancreatic arteries and veins. We use the three-dimension full resolution training strategy which is randomly cropping each patch with the fixed size as the input to the U-shape segmentation network. The patch size is calculated to be $80 \times 192 \times 160$ in both the artery and vein segmentation experiments. The Cross-Entropy (CE) loss and Dice loss are used with 1:1 weight in the training process. The sliding window strategy is adopted for testing.

\subsection{Iterative Trunk Growth Module (ITGM)}

We propose an iterative trunk growth module (ITGM) shown in the blue dashed box in Figure \ref{pipeline} to obtain more complete and fully connected vein predictions based on the basic segmentation results. Our proposed ITGM is composed of a series of trunk growth modules (TGMs) shown in the green dashed box in Figure \ref{pipeline}. The TGM firstly obtains the trunk of the previous segmentation by the largest connected constraint (LCC) operation which only chooses the largest connected component of the vessel prediction, considering the major connected component is the most reliable prediction which is with little false positive (FP) predictions and much false negative (FN) predictions. Then, a branch proposal network (BPN) is designed to find possible missing branches by learning the similar vein's deep features. Although these proposals generated by BPN cannot be absolutely true prediction, the predicted branches which are connected to the raw trunk have high confidence to be the missing branches. Therefore, another LCC operation is adopted for removing the branches which are far way from the previous trunk to obtain a high-quality branches to make the trunk prediction grow. And the above trunk growth process can be iterative to get more complete and fully connected vein prediction. We formulate the iterative trunk growth mechanism as follows:

\begin{equation}\label{trunk}
  Trunk_{i+1}=LCC(BPN(Trunk_{i};Image)+Trunk_{i}).
\end{equation}

\begin{figure}[htb]
		\centering
		{
			\includegraphics[width=9cm,]{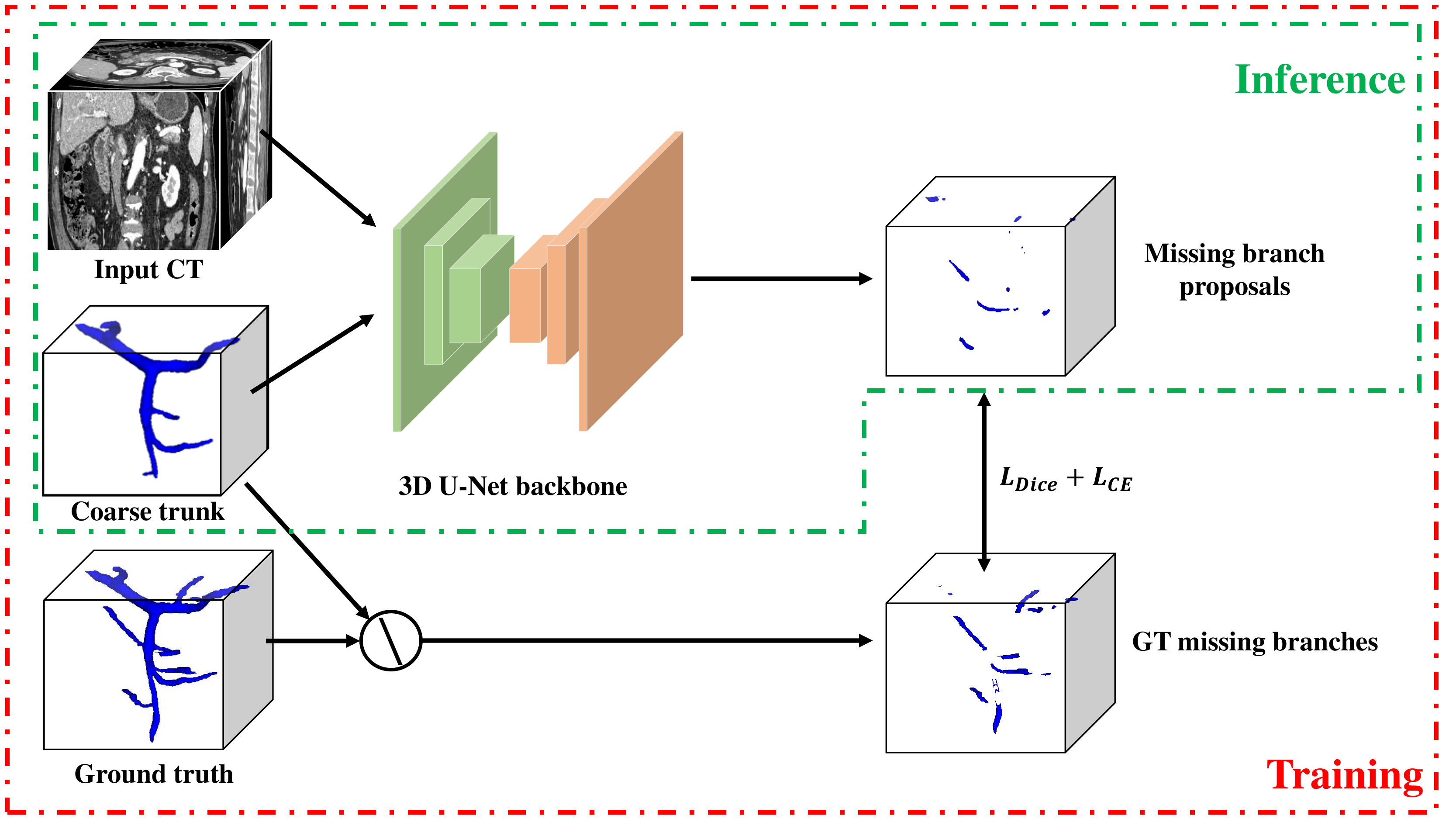}
		}
		\caption{Details of our proposed Branch Proposal Network (BPN).}\label{BPN}
	\end{figure}

The proposed BPN is designed to find the possible false negative regions or missing branches which should be added to the previous vessel trunk. The details of the BPN is shown in Figure \ref{BPN}. When we get the previous vessel trunk, we get the ground truth (GT) of the missing branches by calculating the difference between vessel GT and predictions. Then, a 3D U-Net is trained to learn the feature and distribution of these missing branches. We use the combination of Dice loss and CE loss to train the 3D U-Net in our experiments. In the testing stage, we apply the trained BPN to the previous trunks to get possible missing branch proposals. It should be pointed that, in our experiments, the BPN is only trained once and iteratively applied to inference on the subsequent vein predictions.

\FloatBarrier

\subsection{Weakly supervised labeling mechanism (WSLM)}

\begin{figure}[htb]
		\centering
		{
			\includegraphics[width=9cm,]{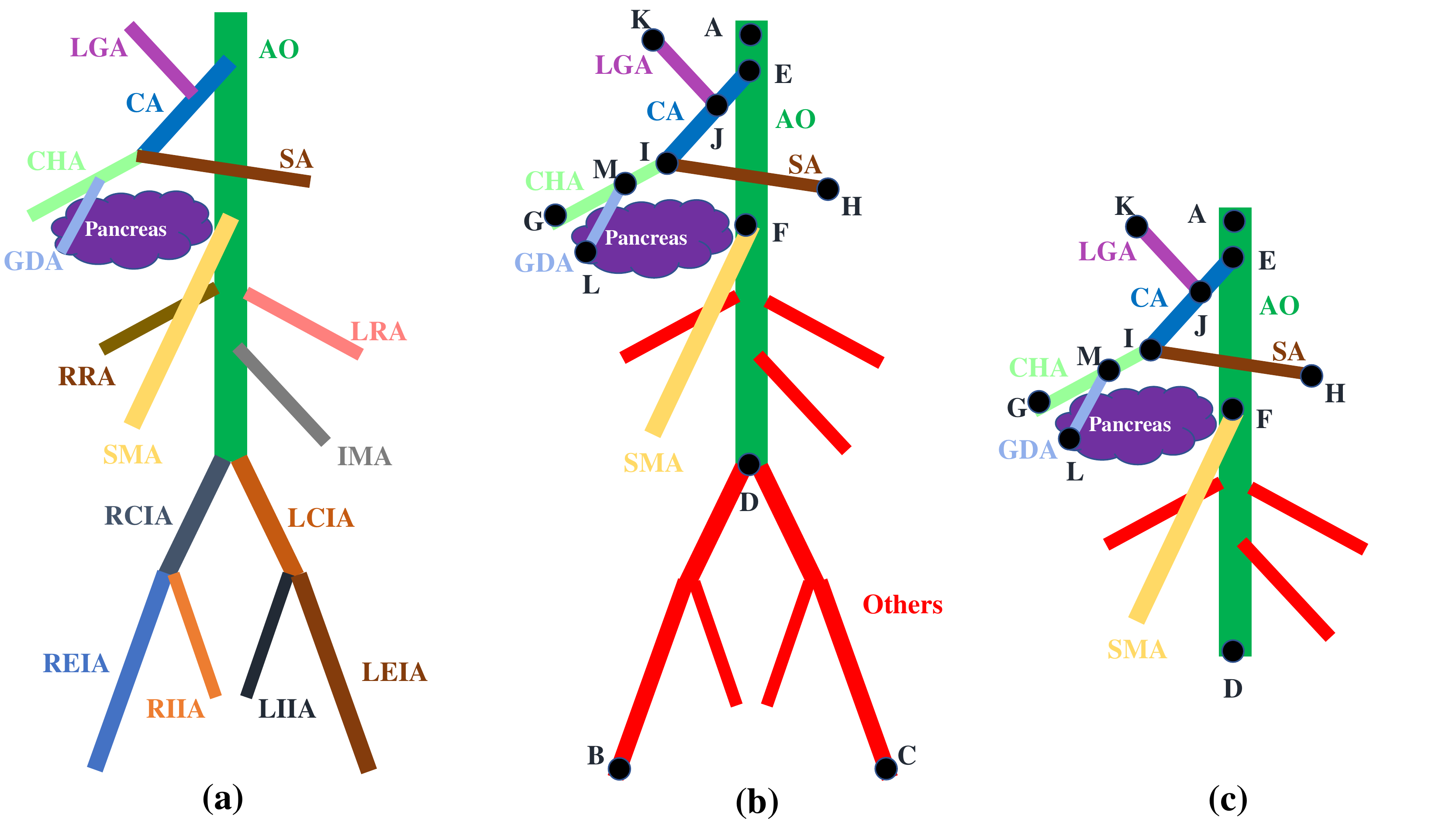}
		}
		\caption{(a) A schematic illustration of the abdominal artery structure. (b) The anatomical labeling problem we address for peripancreatic arteries in this work. (c) An example of the upper abdominal peripancreatic artery structure.}\label{rule}
	\end{figure}

The structure of the abdominal arteries is shown in Figure \ref{rule}(a). We aims to automatically identify the seven branches including AO, CA, SMA, SA, CHA, LGA and GDA, which are most important for pancreatic tumor diagnosis and treatment. Figure \ref{rule}(b) shows the above branches and related keypoints. Besides, we also consider the upper abdominal vascular system without the iliac arteries which is shown in Figure \ref{rule}(c).

\FloatBarrier

\subsubsection{Rule-based preprocessing}
We extract the peripancreatic artery volumes based on the prediction of the basic segmentation network. We skeletonize the basic artery volumes by the method proposed in \cite{Lee}. Then, we build the graphs on these skeletons and get the keypoints of these skeletons by neighbor selection. Finally, we get the vessel tree structures by the Prim algorithm.

After we get the tree structure of the peripancreatic artery, AO is the first target branch we are going to label. AO passes through the center of  the abdominal region in a human body and bifurcates into iliac arteries in the lower abdominal region. By utilizing the anatomical knowledge, anatomical labeling of AO is performed as follows. For the whole abdominal CT case shown in Figure 6(b), we find the most head side endpoint $A$, and two endpoints $B$ and $C$ which have larger diameter than others at the most foot side. We denote $A \rightarrow B$ as the shortest path from $A$ to $B$ in the tree structure. By get the common part of $A \rightarrow B$ and $A \rightarrow C$, then we get the conjunction $D$, and AO is exactly labeled as $A \rightarrow D$. As for the upper abdominal case shown in Figure 6(c), we find the most head side endpoint $A$ and the most foot side endpoint $D$, then we can directly label AO as $A \rightarrow D$. We judge that there are iliac arteries or not by comparing the radius between the head side and the foot side.

After we label AO, we can get the junctions on AO centerline. The two most head side junctions, $E$ and $F$, are the CA and SMA junction respectively. By traveling all the endpoints to these two junctions, each path which does not coincide with AO is labeled as CAs (CA and its subordinate branches) and SMA, respectively.

Then, we find the most right and left side endpoints $G$ and $H$ on CAs. $E  \rightarrow I$ is labeled as CA which is the common path of $E \rightarrow G$ and $E \rightarrow H$. $I \rightarrow H$ is labeled as SA and $I \rightarrow G$ is labeled as CHA. For LGA, we have

\begin{equation}\label{LGA}
	LGA=(\underset{i \in U_{CAs}}{\arg\min} |(i \rightarrow E) \cap CA |) \rightarrow E,
\end{equation}
where $U_{CAs}$ is the set of endpoints on CAs. Finally, The rest of the CAs $M \rightarrow L$ is labeled as GDA.

\subsubsection{Anatomical labeling network}
\begin{figure}[htb]
		\centering
		{
			\includegraphics[width=9cm,]{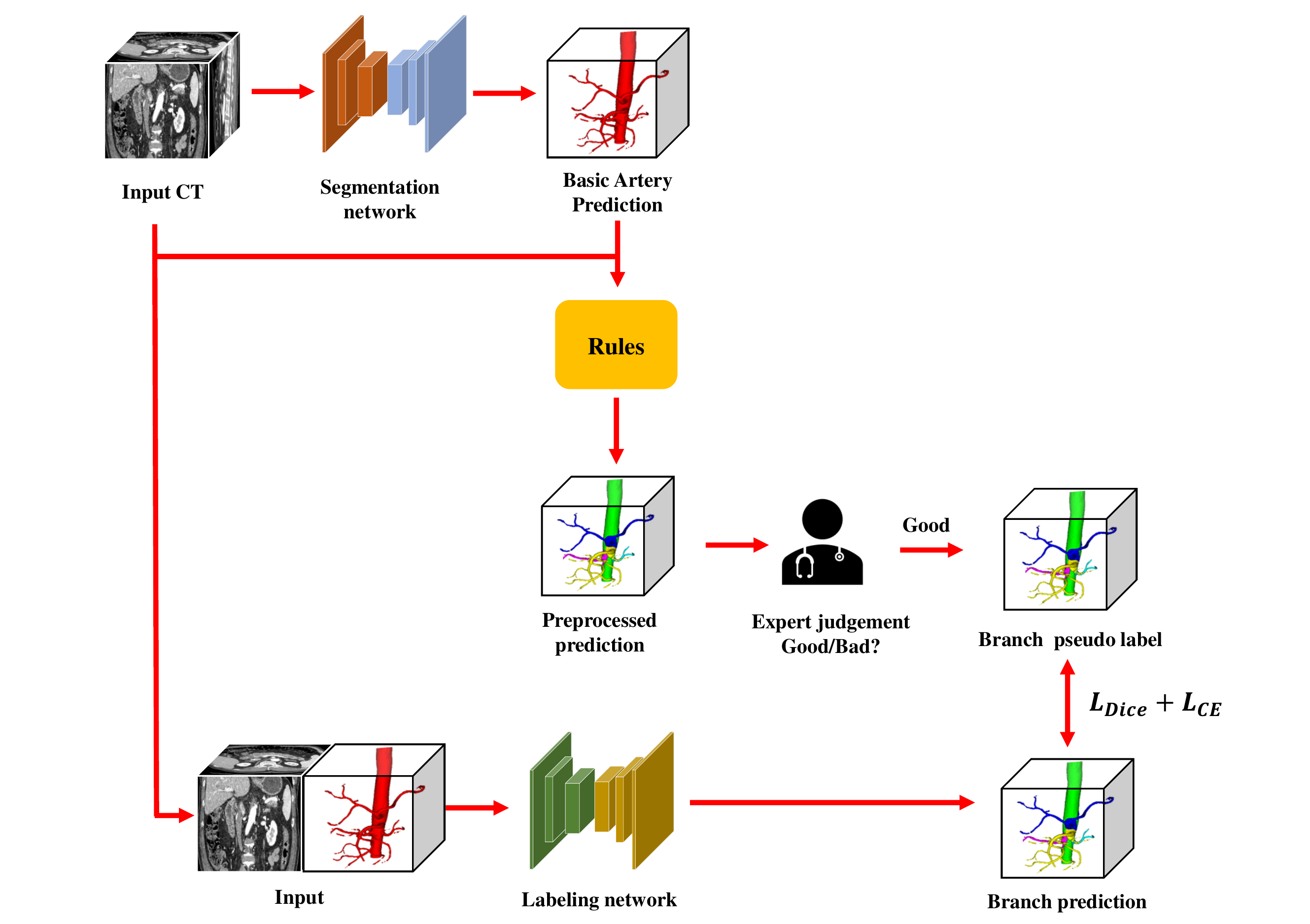}
		}
		\caption{Details of our proposed Weakly Supervised Labeling Mechanism (WSLM).}\label{wslm}
	\end{figure}

After we obtain the primary artery labeling results by the above rule-based strategy, the experts only give the patient-level annotation of ``good'' or ``bad'' for the preprocessed predictions. The ``good'' judgement is given when the true positive voxels account for the majority on all focused branches and there is no negative influence for clinical issues based on these predictions. The ``good'' predictions are used as pseudo branch labels for subsequent procedures. Then we train an atomical labeling network based on the 3D U-Net architecture by feeding the original CT image and basic vessel prediction as inputs, and calculating the Dice and CE loss between the predictions and pseudo branch labels voxel by voxel. The details of the whole WSLM technique is shown in Figure \ref{wslm}.

\FloatBarrier

\subsubsection{Adaptive radius-based post-processing}
\begin{figure}[htb]
		\centering
		{
			\includegraphics[width=8cm,]{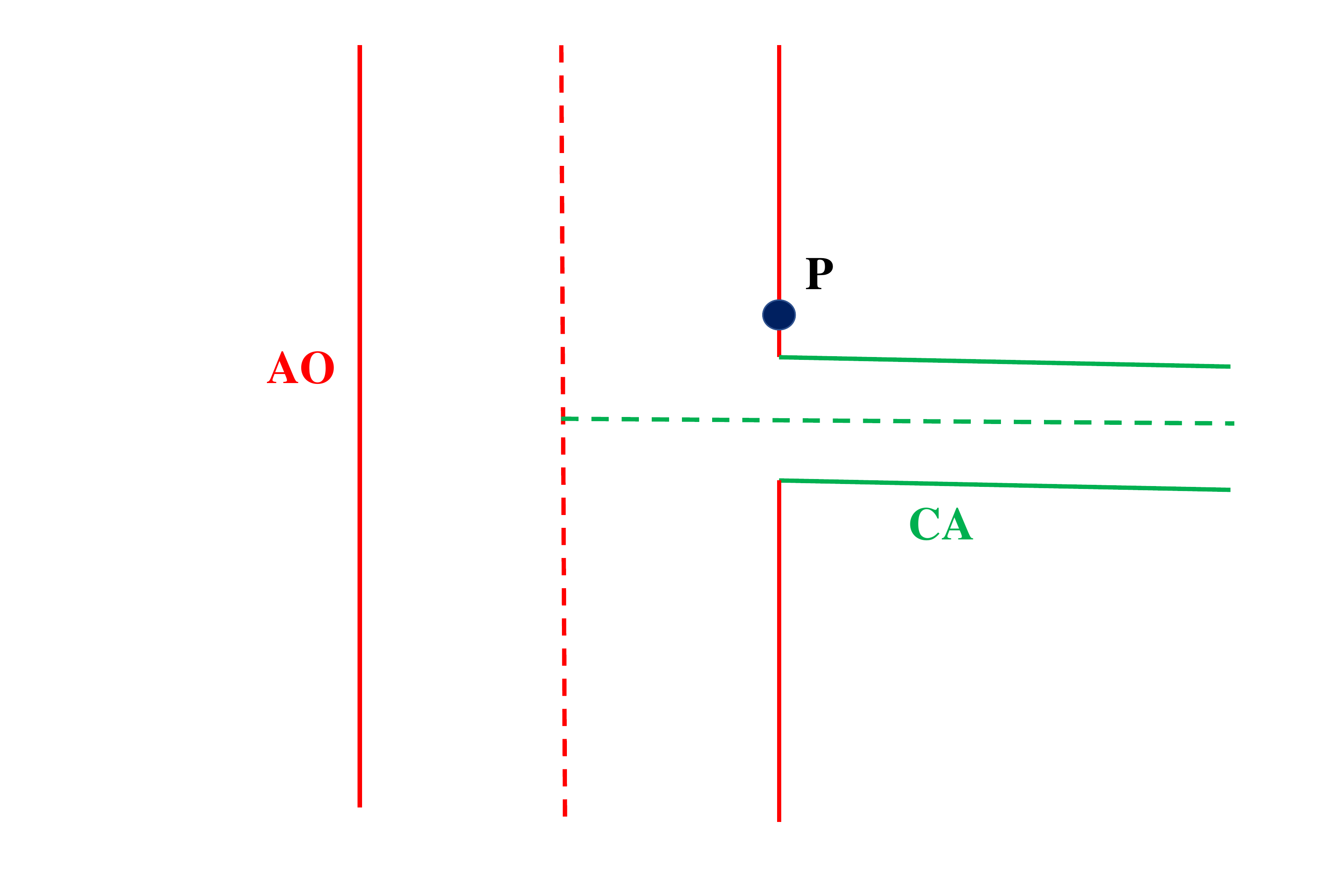}
		}
		\caption{A schematic diagram for the reason that mapping from Eq. (3) predicts ``disk'' branch results. The dash lines denote the centerlines of the branches.}\label{post1}
	\end{figure}

We observe that the rule-based preprocessing brings ``disk'' branch annotations to the subsequent labeling network because of the large gap between AO's radius and other branches', which makes negative impact on labeling evaluation for artery branches especially CA.  A schematic diagram for this happening is shown in Figure \ref{post1}. After labeling the points on the vessel tree, we build a mapping from the centerline to the 3D volume:

\begin{equation}\label{cltovol}
	K(p)=K(\underset{i \in U_{cl}}{\arg\min} ||i-p||_{2}),
\end{equation}
where $p$ is a point on the 3D volume, $K(p)$ is the branch prediction of $p$, $U_{cl}$ is the set of points on the centerline. In Figure \ref{post1}, the distance from point p to AO's centerline is farther than CA's. Therefore, point $p$ and its surrounding voxels will be wrongly predicted to CA branch category.

\begin{figure}[htb]
		\centering
		{
			\includegraphics[width=9cm,]{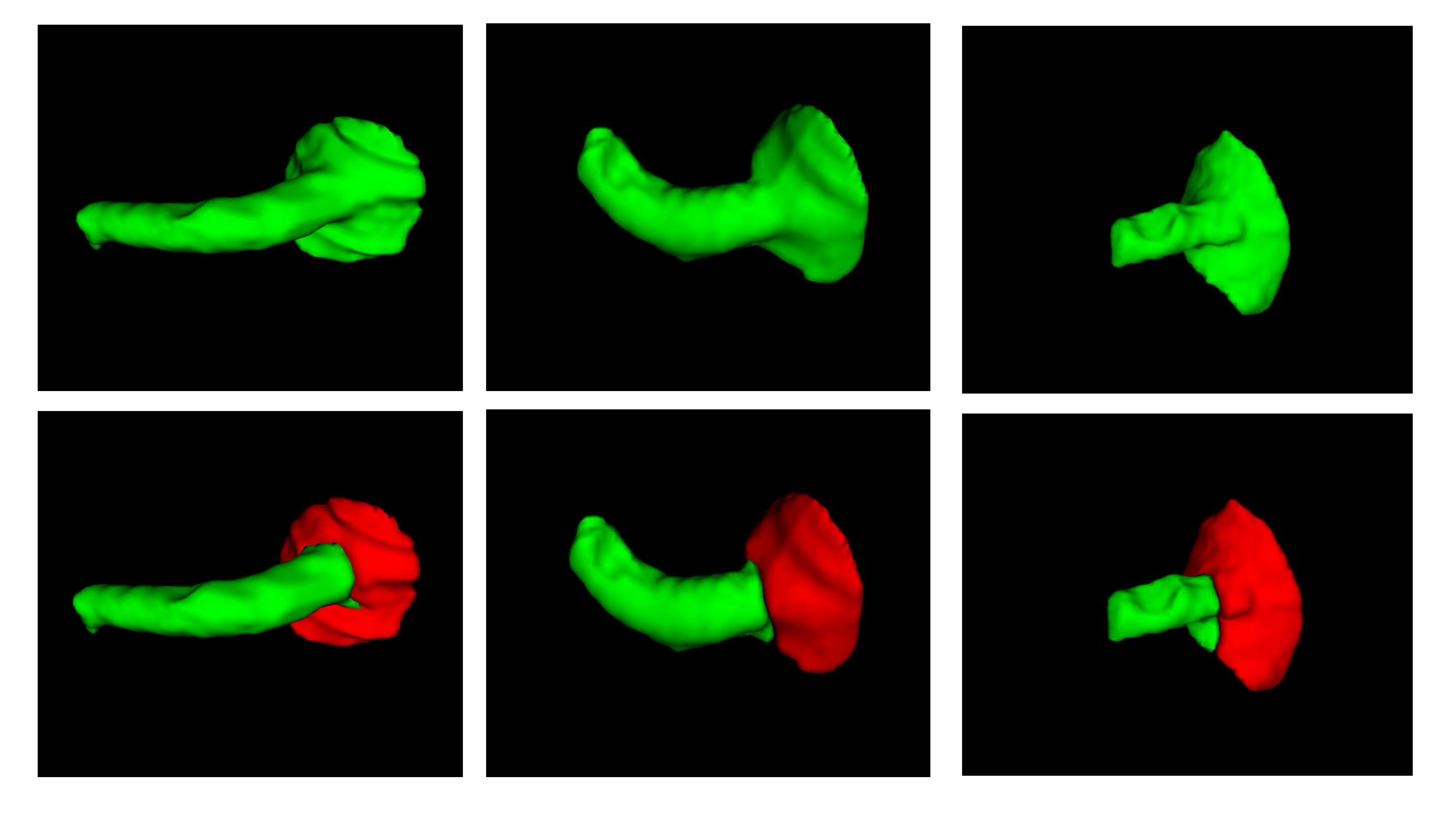}
		}
		\caption{Visual comparison of the labeling results with (the second row) and without (the first row) the adaptive radius-based post-processing. Red and green denote AO and CA predictions, respectively.}\label{post2}
	\end{figure}

To solve this problem, we design an adaptive radius-based post-processing. Considering the wrong ``disk'' predicted volumes have larger radius, we calculate the radius of each point on the branch centerline. Then, we use K-Means algorithm to divide these points into two clusters. The cluster with larger radius is predicted as AO. Figure \ref{post2} shows the visual comparison of the labeling results with and without our adaptive radius-based post-processing. It should be pointed that, because CA has smaller size and is more affected by the ``disk'' predictions than the other branches, we only apply our proposed adaptive radius-based post-processing to CA in our experiments. The other affected branches such as SMA can be post-processed in the same mechanism.

\FloatBarrier
\FloatBarrier
\section{Experiments}
\subsection{Dataset and evaluation}

In this work, we evaluate our proposed method on two datasets collected from Nanjing Drum Tower Hospital. The first dataset, called peripancreatic vein segmentation (PPV) dataset, consists of 272 venous phase CT scans from patients with surgical pathology-confirmed pancreatic tumors (136 PDACs, 43 IPMNs, 43 SCNs, 24 SPTs, 26 MCNs). The second dataset, called peripancreatic artery segmentation (PPA) dataset, consists of 338 arterial phase CT scans from patients with surgical pathology-confirmed pancreatic tumors (191 PDACs, 46 IPMNs, 40 SCNs, 34 SPTs, 27 MCNs). Each CT scan of PPV or PPA has a manual annotation of peripancreatic vein or artery respectively which is performed by the pancreatic imaging radiologists. All the segmentation experiments are performed using nested five-fold cross-validation.

We employ the Dice similarity coefficient as the segmentation metrics which can be calculated as follows.
\begin{equation}\label{dsc}
  Dice=\frac{2 \times TP}{2 \times TP+FP+FN},
\end{equation}
where $TP$, $FP$ and $FN$ in Eq. (4) denote the numbers of true positive, false positive and false negative voxels in the prediction, respectively. In addition to the above traditional metrics, we adopt the following two segmentation metrics for distractions. We use the false positive segmentation rate (FPSR) and false negative segmentation rate (FNSR), which can intuitively reflect the false positive and false negative segmentation errors. They can be calculated as follows.
\begin{equation}\label{fpsr}
  FPSR=\frac{FP}{2 \times TP+FP+FN},
\end{equation}

\begin{equation}\label{fnsr}
  FNSR=\frac{FN}{2 \times TP+FP+FN}.
\end{equation}
We also employ the clDice \cite{cldice} to measure the topology preservation of the segmentation for targets with tubular structures. The clDice can be calculated as follows.

\begin{equation}\label{tprec}
  Tprec(S_{P},V_{G})=\frac{|S_{P} \cap V_{G}|}{|S_{P}|},
\end{equation}

\begin{equation}\label{tsens}
  Tsens(S_{G},V_{P})=\frac{|R_{G} \cap V_{P}|}{|S_{G}|},
\end{equation}

\begin{equation}\label{cldice}
  clDice(V_{P},V_{G})=2 \times \frac{Tprec \times Tsens}{Tprec+Tsens},
\end{equation}
where $V_{G}$ and $V_{P}$ are the ground truth mask and the predicted segmentation mask, respectively. $S_{G}$ and $S_{P}$ are the skeletons extracted from $V_{G}$ and $V_{P}$, respectively.

To evaluate the connectivity-preserving performance of segmentation, we also calculate the number of connected components (NCC) in our experiments.

Branch-wise performance for anatomical labeling is evaluated by the recall rate, precision rate and F-measure (or $F_1$ measure) which are calculated as follows.

\begin{equation}\label{precision}
  Precision=\frac{TP}{TP+FP}
\end{equation}

\begin{equation}\label{recall}
  Recall=\frac{TP}{TP+FN}
\end{equation}

\begin{equation}\label{F1}
  F_1=2 \times \frac{Recall \times Precision}{Precision + Recall}
\end{equation}

We define a TP branch prediction while TP voxels on this branch account for the majority compared with FP and FN voxels. FP and FN branch can be defined in a similar way.

\subsection{Vein segmentation}

\subsubsection{Vein segmentation results on PPV dataset}
We get the peripancreatic vein segmentation results of our proposed APESA for different pancreatic tumor cases in our collected PPV dataset through five-fold cross-validation experiments. We show the metrics of the clDice, Dice, FPSR and FNSR in Table \ref{tabdifftumor}. The highest metric is shown in bold and the lowest one is shown in red. As we can see, the non-PDAC cases achieve better segmentation performance comparing to the PDAC cases, which may be caused by the invasion of peripancreatic vessels by pancreatic malignant tumors.

\begin{table}[htb]
        \centering
		\caption{Comparison of segmentation results for pancreatic veins with different pancreatic tumors on our PPV dataset.}
		\vspace{0pt}	
		\renewcommand\arraystretch{2}
		\setlength{\tabcolsep}{6mm}{}{
			\begin{tabular}{c|c|c|c|c}
				\hline
				Case   &clDice($\%\uparrow$)  & Dice($\%\uparrow$)   &FPSR($\%\downarrow$)  & FNSR($\%\downarrow$)    \\
				\hline
				PDAC   &\textcolor{red}{84.17$\pm$14.68}       &\textcolor{red}{91.44$\pm$11.16}      &\textcolor{red}{5.66$\pm$9.70}          &\textcolor{red}{2.90$\pm$2.92}\\
                IPMN   &89.75$\pm$11.01       &95.32$\pm$5.89        &2.94$\pm$5.77          &1.74$\pm$0.90\\
                MCN    &92.51$\pm$6.67        &97.16$\pm$3.73       &0.83$\pm$3.59         &2.02$\pm$1.34\\
                SCN    &\textbf{94.14$\pm$5.26}  &\textbf{97.73$\pm$2.75}       &\textbf{0.36$\pm$2.14}           &1.91$\pm$1.83\\
                SPT    &90.34$\pm$13.19        &96.24$\pm$5.42       &2.41$\pm$5.37          &\textbf{1.36$\pm$0.69} \\
                \hline
                Total     &87.96$\pm$12.93   &94.01$\pm$8.96 &3.65$\pm$7.85  &2.34$\pm$2.34 \\
				\hline
		\end{tabular}}
		\label{tabdifftumor}
	\end{table}

\subsubsection{Ablation study}
\begin{figure}[htb]
		\centering
		{
			\includegraphics[width=9cm,]{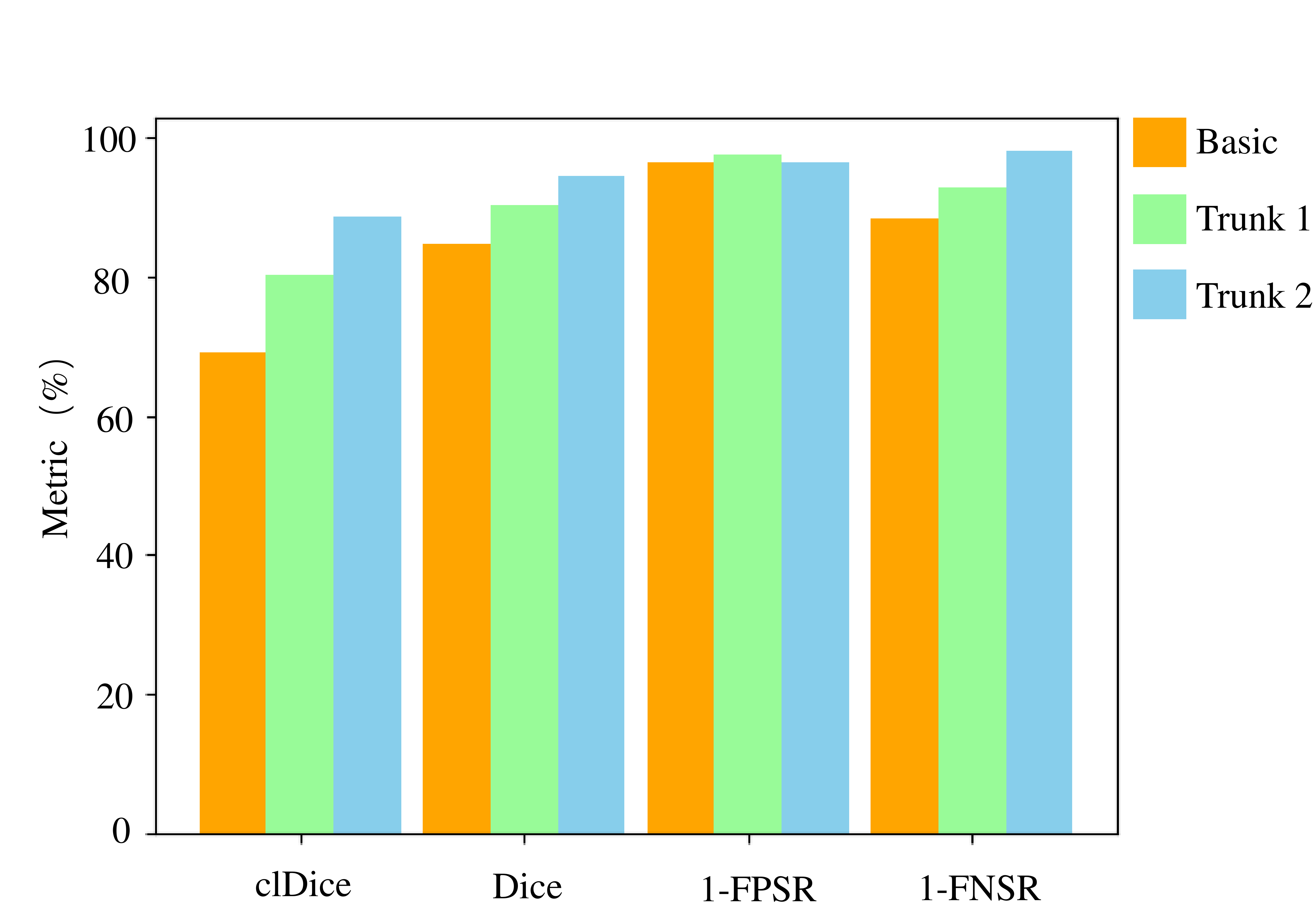}
		}
		\caption{Visual comparison results of the ablation study in terms of different metrics.}\label{abbar}
	\end{figure}

\begin{figure}[htb]
		\centering
		{
			\includegraphics[width=9cm,]{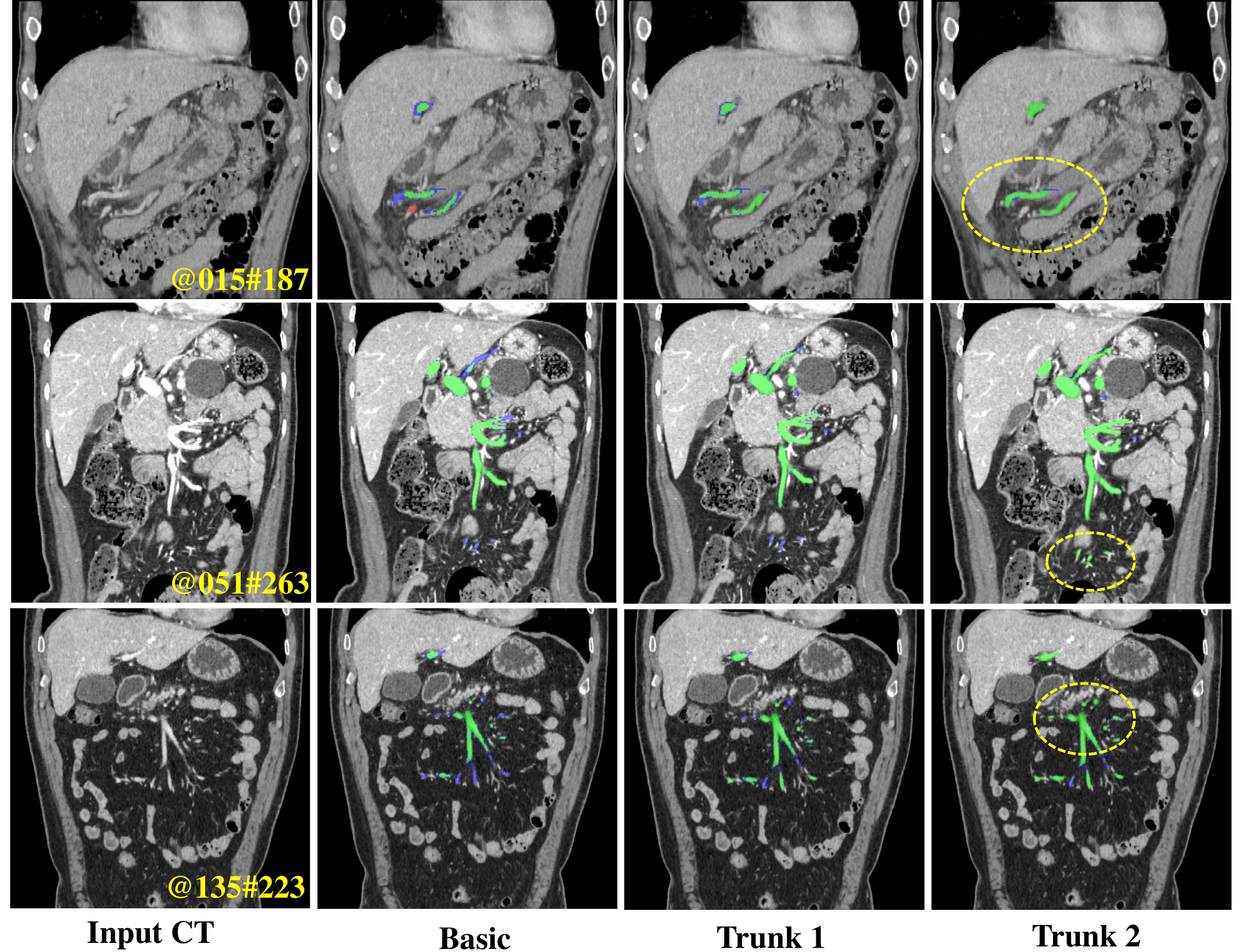}
		}
		\caption{Visual comparison results of the ablation study. The green, red and blue pixels denote the true positive, false positive and false negative segmentation, respectively.}\label{abvis2d}
	\end{figure}

\begin{figure}[htb]
		\centering
		{
			\includegraphics[width=9cm,]{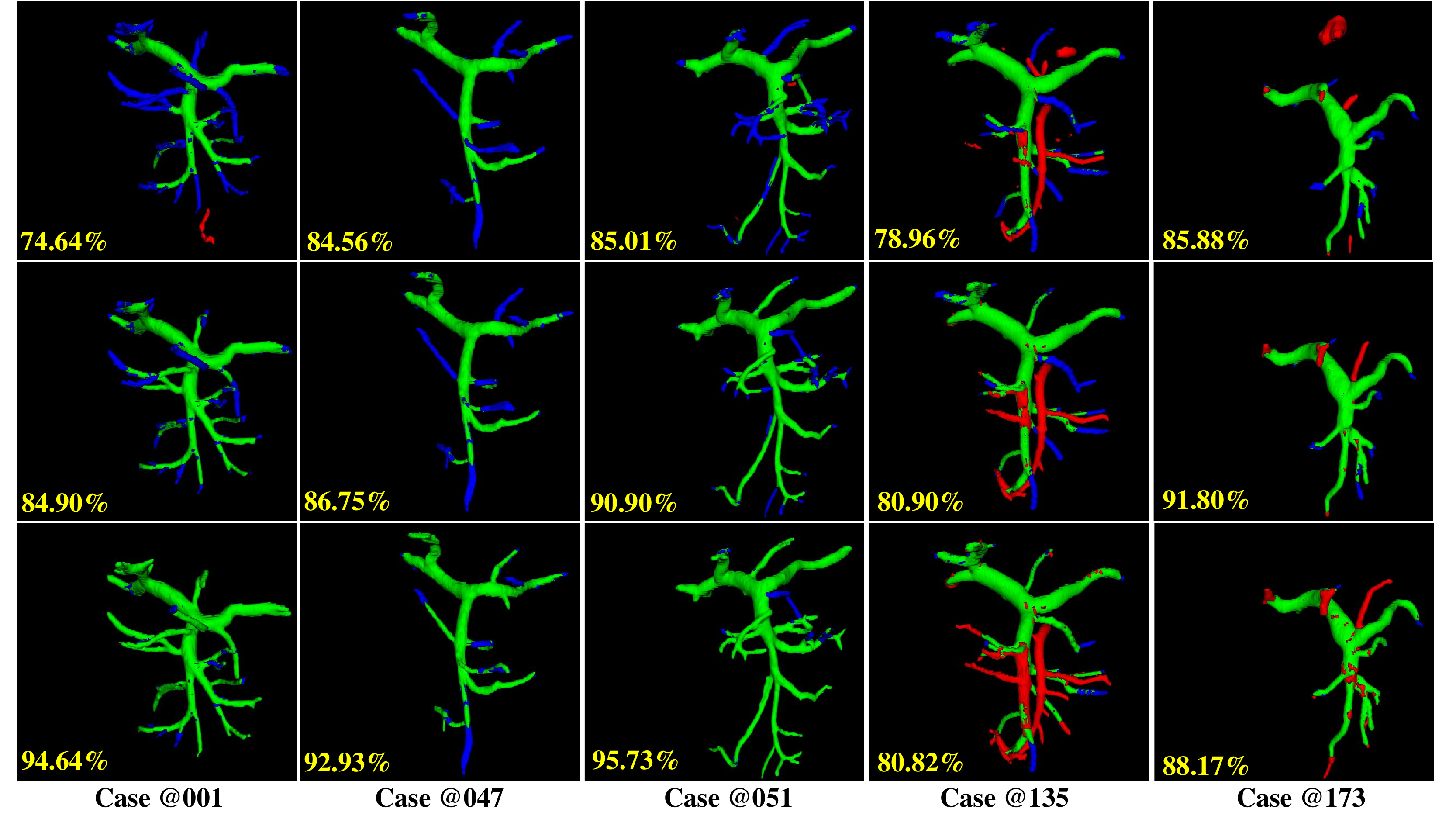}
		}
		\caption{The 3D render comparison results of the ablation study. The green, red and blue voxels denote the true positive, false positive and false negative segmentation, respectively. Dice score of each case is written in the bottom left of each image. \textbf{Top:} basic segmentation. \textbf{Middle:} fine segmentation. \textbf{Bottom:} refined segmentation. }\label{abvis3d}
	\end{figure}

To verify the improvement of our proposed ITGM, we conduct the ablation study results shown in Table \ref{tabab}. We compare the segmentation performance of the nnU-Net in the basic segmentation stage marked as basic, and the two following TGM results after LCC marked as trunk 1 and trunk 2. Besides, we also list the performance of largest connected component of the basic prediction marked as trunk 0, and the iterative grown predictions before LCC operation which are marked as TGM $\times 1$ and TGM $\times 2$, respectively. From Table II, we can see that the proposed ITGM can effectively boost the topological integrity (clDice) and the overall segmentation (Dice) performance. Additionally, the proposed method can significantly improve the false negative predictions and keep comparable false positive segmentation performance to the basic predictions. We can also see that the ITGM can completely ensure the connectivity, and has better performance than the largest connected component of the basic predictions. Figure \ref{abbar} presents the performance distributions of different stages. We also show the visualization of the different segmentation results in Figure \ref{abvis2d} (coronal plane) and Figure \ref{abvis3d} (3D render).

\begin{table}[htb]
        \centering
		\caption{Ablation study for peripancreatic vein segmentation on our proposed PPV dataset.}
		\vspace{0pt}	
		\renewcommand\arraystretch{2}
		\setlength{\tabcolsep}{4mm}{}{
			\begin{tabular}{c|c|c|c|c|c}
				\hline
			    Methods&clDice($\%\uparrow$)&Dice($\%\uparrow$)&FPSR($\%\downarrow$) &FNSR($\%\downarrow$) &NCC($\downarrow$)\\
				\hline
				Basic & 68.52$\pm$9.22 &84.57$\pm$7.90 & 3.43$\pm$8.54 &12.00$\pm$4.31 &5.08$\pm$2.72\\
				Trunk 0   & 68.66$\pm$8.76 & 85.52$\pm$6.86 & \textbf{1.44$\pm$5.28} & 13.04$\pm$4.89   &\textbf{1.00$\pm$0.00}\\
                TGM $\times 1$   & 78.68$\pm$9.75 & 89.53$\pm$7.33 & 3.16$\pm$6.73 & 7.31$\pm$3.31  &4.33$\pm$2.29\\
                Trunk 1   & 79.57$\pm$9.15 & 89.98$\pm$7.04 & 2.46$\pm$6.31 & 7.56$\pm$3.53  &\textbf{1.00$\pm$0.00} \\
                TGM $\times 2$   & 86.12$\pm$12.79 & 93.41$\pm$8.53 & 4.45$\pm$7.99 & \textbf{2.13$\pm$1.76}  &6.26$\pm$3.32 \\
                Trunk 2     & \textbf{87.96$\pm$12.93} & \textbf{94.01$\pm$8.96}  & 3.65$\pm$7.85          &2.34$\pm$2.34        &\textbf{1.00$\pm$0.00} \\
				\hline
		\end{tabular}}
		\label{tabab}
	\end{table}

\FloatBarrier
\FloatBarrier
\subsubsection{Comparison to SOTA segmentation methods}

We also compare to other baseline models for peripancreatic vein segmentation in our PPV dataset. Recently, Transformer \cite{transformer}, a sequence-to-sequence prediction framework, has been considered as an alternative architecture, and has achieved competitive performance on many computer vision tasks, like image recognition \cite{vit}, semantic segmentation \cite{swin}, object detection \cite{carion} and low-level vision \cite{parmar}.  CoTr \cite{cotr} and UNETR \cite{unetr} are the combination of convolutional neural network and transformer for computer vision and achieve state-of-the-art (SOTA) in medical image segmentation challenges. We compare above two SOTA methods and the nnU-Net \cite{nnunet} to our proposed APESA on our PPV dataset. All the experiments follow the same 5-fold cross-validation setting.

\begin{figure}[htb]
		\centering
		{
			\includegraphics[width=9cm,]{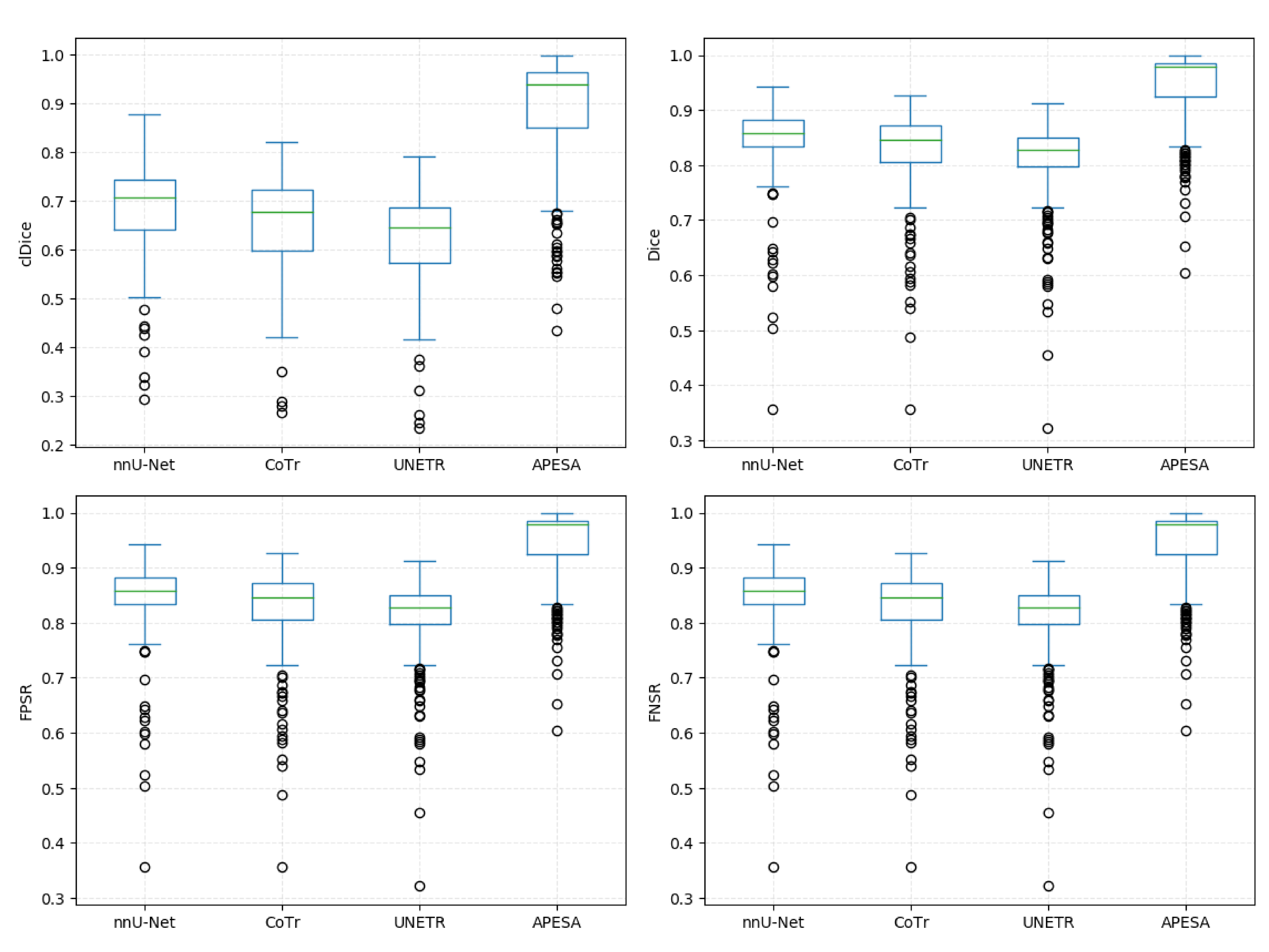}
		}
		\caption{Performance distributions of different methods for peripancreatic vein segmentation on PPV datasets. }\label{cpbox}
	\end{figure}

\begin{figure}[htb]
		\centering
		{
			\includegraphics[width=9cm,]{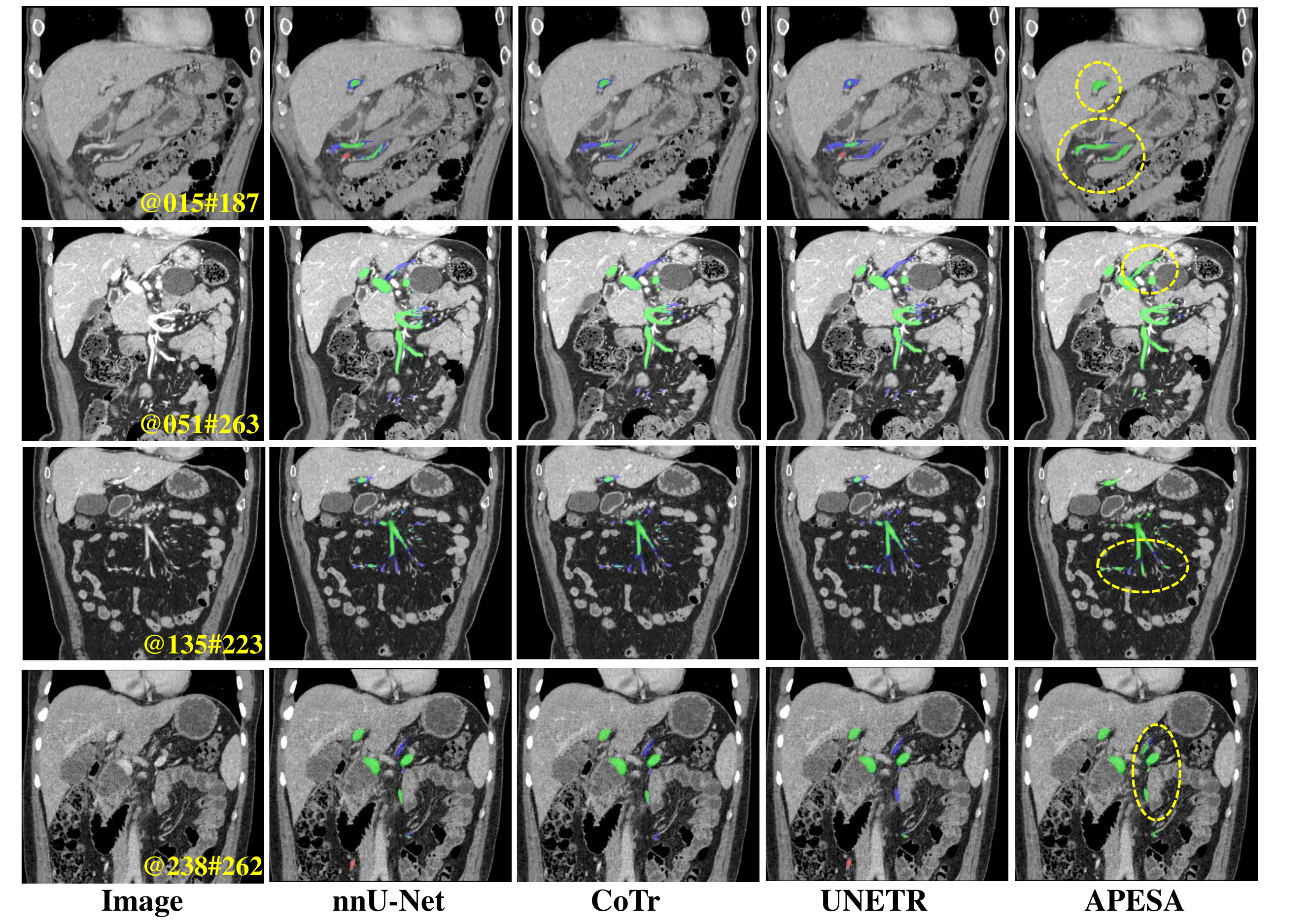}
		}
		\caption{Visual comparison results to other baseline methods. the green, red and blue contours denote the ground truth and the predicted segmentation, respectively.}\label{cpvis2d}
	\end{figure}

\begin{figure}[htb]
		\centering
		{
			\includegraphics[width=9cm,]{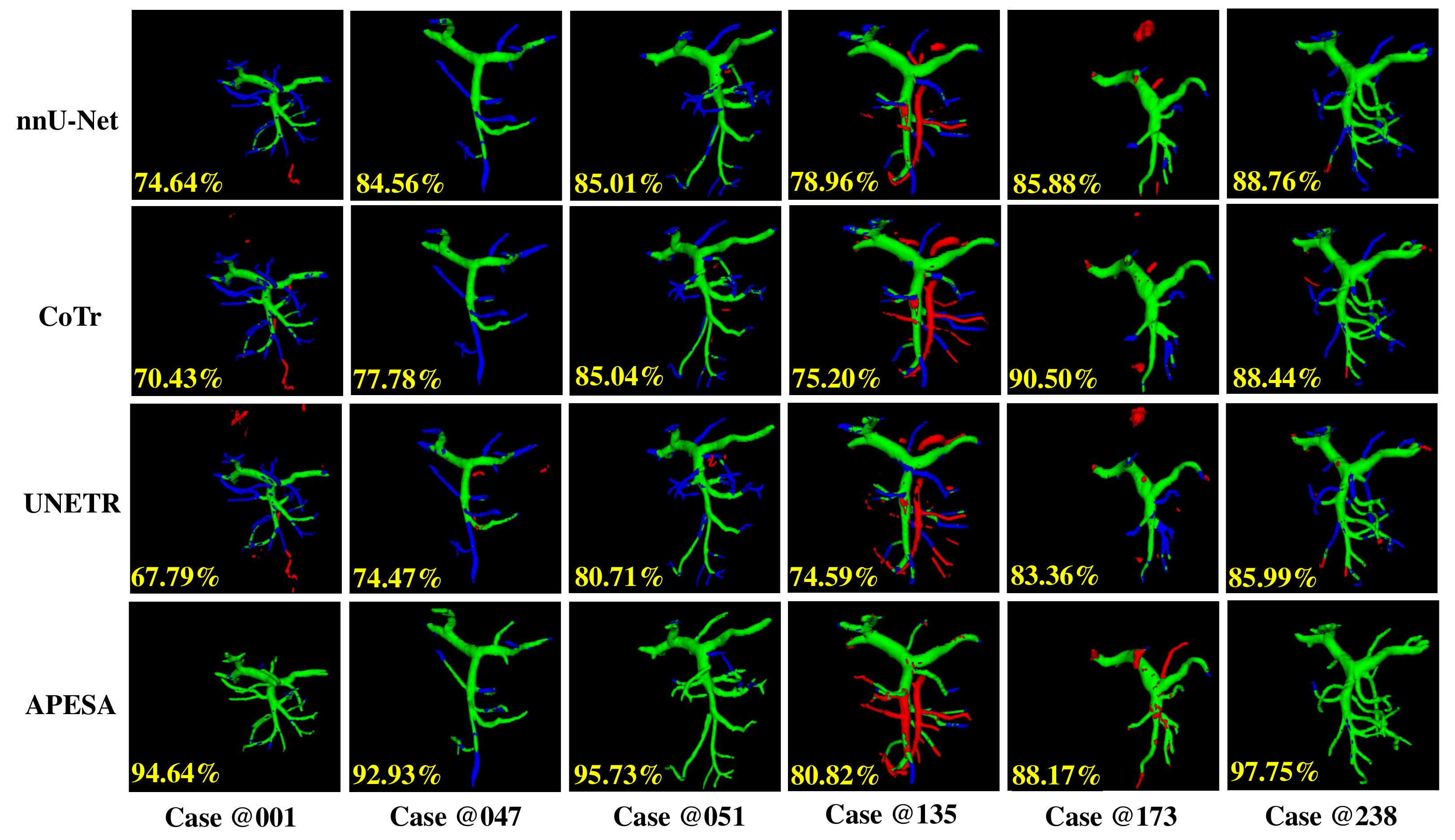}
		}
		\caption{The 3D render comparison results to other baseline methods. the green, red and blue voxels denote the true positive, false positive and false negative segmentation, respectively. Dice score of each case is written in the bottom left of each image. }\label{cpvis3d}
	\end{figure}

The segmentation results are shown in Table \ref{tabcp} where we can see that our APESA achieves the best performance in terms of clDice, Dice, FNSR and NCC metrics. Additionally, the transformer-combined methods do not show their superiority to the CNN-based method, which may be caused by the limited scale of the dataset. Figure \ref{cpbox} presents the performance distributions of different methods. It can be found that our proposed APESA has a more compact distribution with fewer outliers. Figure \ref{cpvis2d} and Figure \ref{cpvis3d} show the visual comparisons of coronal plane and 3D render results, respectively.
\begin{table}[htb]
        \centering
		\caption{Comparison to SOTA segmentation methods for peripancreatic vein segmentation on PPV dataset.}
		\vspace{0pt}	
		\renewcommand\arraystretch{2}
		\setlength{\tabcolsep}{5mm}{}{
			\begin{tabular}{c|c|c|c|c|c}
				\hline
			    Methods&clDice($\%\uparrow$)&Dice($\%\uparrow$)&FPSR($\%\downarrow$) &FNSR($\%\downarrow$)&NCC($\%\downarrow$)\\
				\hline
                nnU-Net & 68.52$\pm$9.22 &84.57$\pm$7.90 & \textbf{3.43$\pm$8.54} &12.00$\pm$4.31 &5.08$\pm$2.72\\
				CoTr &65.30$\pm$10.10   &82.47$\pm$8.56 &4.00$\pm$7.17  &13.52$\pm$6.82 &6.64$\pm$3.56 \\
                UNETR &62.25$\pm$10.21   &80.90$\pm$8.69 &5.42$\pm$9.24  &13.68$\pm$4.70  &15.13$\pm$7.36\\
                \hline
                \textbf{APESA}   &\textbf{87.96$\pm$12.93} & \textbf{94.01$\pm$8.96}  & 3.65$\pm$7.85   &\textbf{2.34$\pm$2.34}  &\textbf{1.00$\pm$0.00}\\
				\hline
		\end{tabular}}
		\label{tabcp}
	\end{table}

\FloatBarrier
\FloatBarrier
\FloatBarrier
\subsubsection{Comparison to the works related to vein segmentation.}
To the best of our knowledge, this is the first work concentrating on the segmentation of peripancreatic veins and there is almost no published segmentation performance we can compare. Therefore, we compare with the state-of-the-art methods for portal vein or other related vein branch segmentation in Table \ref{tabsota}. Because any direct comparison is not possible, we list the reported Dice score in these studies. We can see that our methods achieve the highest reported Dice score on abdominal vein segmentation. Besides, since our dataset includes patients with 5 types of pancreatic tumors, our proposed method has more reliable performance for clinical diagnosis and treatment.

\begin{table}[htb]
        \centering
		\caption{Comparison to the works related to vein segmentation.}
		\vspace{0pt}	
		\renewcommand\arraystretch{2}
		\setlength{\tabcolsep}{5mm}{}{
			\begin{tabular}{c|c|c|c|c}
				\hline
				Methods                & Data    &Target & Tumor  & Dice(\%)  \\
				\hline
                Ibragimov 2017 \cite{Ibragimov}   & 72   &Portal vein   & -            &  83.00     \\
                Golla 2020 \cite{Golla}   & 20   &Portal vein   & -            &  75.80     \\
                Mahmoudi 2022 \cite{Mahmoudi}    & 138           &SMV         & 1              & 73.00  \\
                \hline
                \textbf{APESA (Ours)}   &273      &Peripancreatic &\textbf{5}    & \textbf{94.01} \\
				\hline
		\end{tabular}}
		\label{tabsota}
	\end{table}

\subsection{Artery segmentation}

\subsubsection{Segmentation results on PPA dataset}

\begin{figure}[htb]
		\centering
		{
			\includegraphics[width=9cm,]{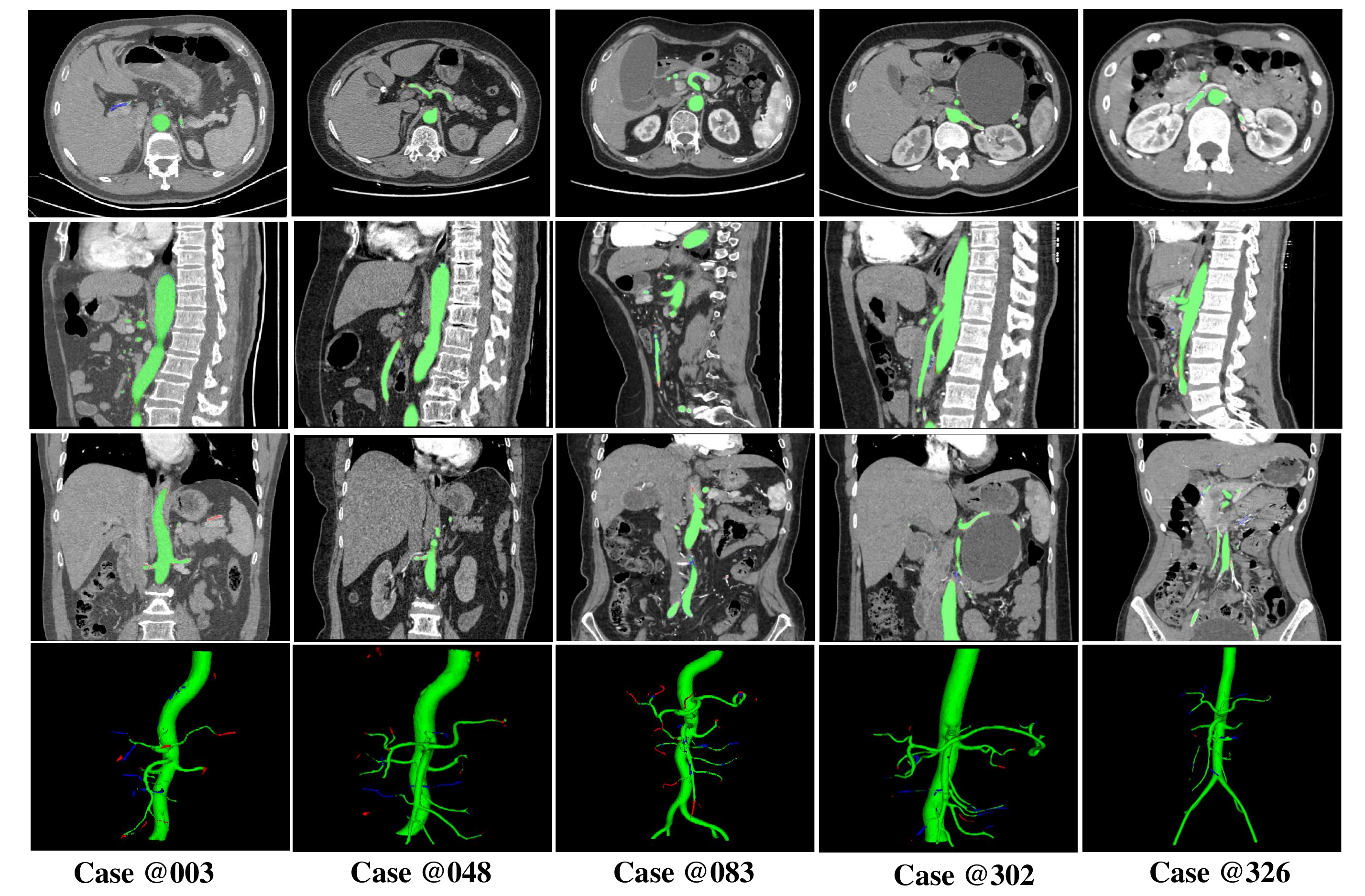}
		}
		\caption{Qualitative results: In the first row our segmentations are presented in the
axial view, in the second row as sagittal view, in the third row as coronal view, in the last row as a 3D rendering. Green, blue, and red represent true positive, false negative and false positive predictions, respectively. }\label{arterysegvis}
	\end{figure}

We also get the peripancreatic artery segmentation results of our proposed APESA for different pancreatic tumor cases in our collected PPA dataset. We show the metrics of the clDice, Dice, FPSR and FNSR in Table \ref{tabarteyseg}. The highest metric is shown in bold and the lowest is shown in red. Just like the vein segmentation performance, the non-PDAC cases achieve better segmentation performance comparing to the PDAC cases. Figure \ref{arterysegvis} presents the qualitative results of the peripancreatic artery segmentation. It can be found that the artery segmentation almost achieves perfect performance comparing to the vein segmentation, which may be caused by the high contrast of arteries in the arterial CT scans.

\begin{table}[htb]
        \centering
		\caption{Comparison of segmentation results for peripancreatic arteries with different pancreatic tumors on our PPA dataset.}
		\vspace{0pt}	
		\renewcommand\arraystretch{2}
		\setlength{\tabcolsep}{6mm}{}{
			\begin{tabular}{c|c|c|c|c}
				\hline
				Case   &clDice($\%\uparrow$)  & Dice($\%\uparrow$)   &FPSR($\%\downarrow$)  & FNSR($\%\downarrow$)    \\
				\hline
				PDAC   &\textcolor{red}{78.98$\pm$8.07}  &\textcolor{red}{96.25$\pm$2.90}  &\textcolor{red}{1.59$\pm$2.40}   &\textcolor{red}{2.16$\pm$1.14}\\
                IPMN   &\textbf{88.15$\pm$3.63}  &\textbf{98.28$\pm$0.34}  &\textbf{0.77$\pm$0.29}   &\textbf{0.95$\pm$0.29}\\
                MCN    &87.13$\pm$5.59  &98.08$\pm$0.66  &0.77$\pm$0.45   &1.15$\pm$0.40\\
                SCN    &85.81$\pm$4.39 &97.85$\pm$0.79  &0.91$\pm$0.63   &1.25$\pm$0.46\\
                SPT    &87.23$\pm$3.13  &97.77$\pm$0.59  &0.95$\pm$0.30   &1.28$\pm$0.43 \\
                \hline
                Total     &82.52$\pm$7.81  &97.01$\pm$2.39 &1.27$\pm$1.87    &1.72$\pm$1.04 \\
				\hline
		\end{tabular}}
		\label{tabarteyseg}
	\end{table}

\FloatBarrier
\subsubsection{Comparison to the works related to peripancreatic artery segmentation}
We also compare with the state-of-the-art methods in peripancreatic artery segmentation in Table \ref{tabarterysegsota}. Dima \textit{et al.} \cite{Dima} get the Dice score of 95.05\% for the peripancreatic artery segmentation using 3D U-Net based on the Iodine and arterial image inputs. Our methods achieve the highest reported Dice score on peripancreatic segmentation by single-phase image input with more tumor types. Additionally, our performance is however no rival for methods using multiple image inputs. Multi-phase CT volumes or Iodine map used in \cite{Dima} can help the model to learn more information about the peripancreatic arteries. However, the annotated dataset with multiple phase images is extremely hard to obtain.

\begin{table}[htb]
        \centering
		\caption{Comparison to the works related to peripancreatic artery segmentation.}
		\vspace{0pt}	
		\renewcommand\arraystretch{2}
		\setlength{\tabcolsep}{4mm}{}{
			\begin{tabular}{c|c|c|c|c}
				\hline
				Methods                           & Data     &Target            & Tumor    & Dice\\
				\hline
                Oda 2019 \cite{Oda2019}    & 20           &Abdominal         & -              &  87.10     \\
                Golla 2020 \cite{Golla}  & 20           &Abdominal         & -              &  83.80     \\
                Dima 2021 \cite{Dima}  & 143          &Peripancreatic    & 1              &  95.05     \\
                Mahmoudi 2022 \cite{Mahmoudi}    & 138           &SMA         & 1              & 81.00  \\
                Zhu 2022 \cite{ZhuOda}    & 20           &Abdominal         & -              & 93.20  \\
                \hline
                \textbf{APESA (Ours)}              &345         &Peripancreatic   &\textbf{5}      & \textbf{97.01} \\
				\hline
		\end{tabular}}
		\label{tabarterysegsota}
	\end{table}

\subsection{Artery branch identification}
\begin{figure}[h]
		\centering
		{
			\includegraphics[width=12cm,]{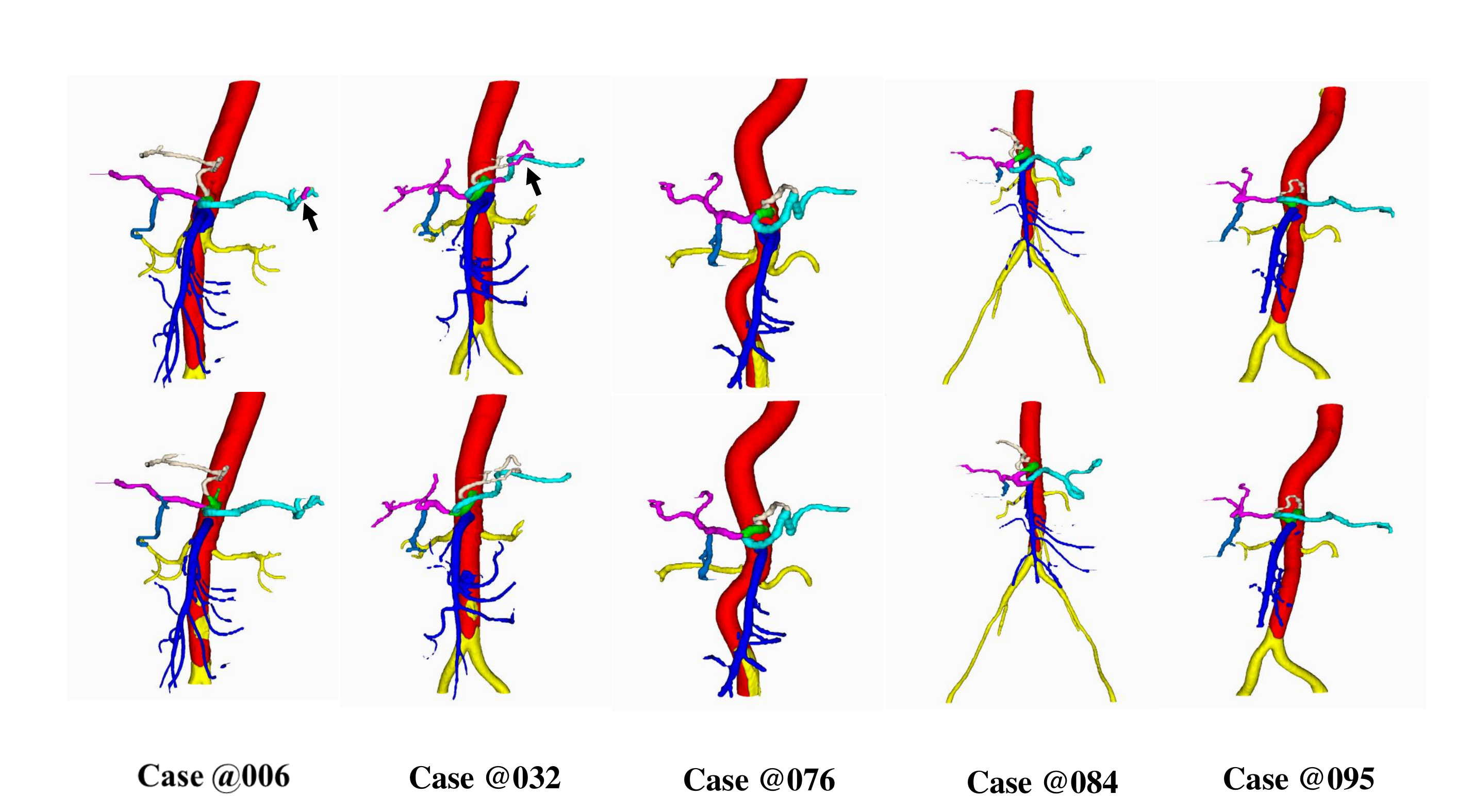}
		}
		\caption{The 3D render comparison results of the anatomical labeling predictions for the peripancreatic artery. \textbf{Top:} Predictions. \textbf{Bottom:} Ground truth. The black arrows show the difference between prediction and ground truth.}\label{parse}
	\end{figure}

We apply the rule-based preprocessing to 108 cases from the PPA dataset. There are 24 cases are judged as ``good'' by experienced experts. We train the labeling network with these 24 pseudo annotations. And there are 30 independent cases from the PPA dataset with voxel-wise branch annotations by experts for testing. Table \ref{tablabeling} shows the anatomical labeling results for the peripancreatic artery branches. Branch-wise and voxel-wise metrics are shown in blue and red, respectively. Since most abdominal artery labeling studies include our peripancreatic artery labeling mission, we list the reported experimental results for the related branches in Table \ref{tabsotalabeling}.  As we can see, we achieve a best or comparable performance for peripancreatic artery branch labeling compared to the existing labeling techniques. Figure \ref{parse} presents the qualitative results of the anatomical labeling predictions, and the difference between predictions and ground truth is marked by arrow.

\begin{table}[htb]
        \centering
		\caption{Experimental results of WSLM for peripancreatic artery branch identification.}
		\vspace{0pt}	
		\renewcommand\arraystretch{2}
		\setlength{\tabcolsep}{6mm}{}{
			\begin{tabular}{c|c|c|c|c}
                \hline
			    Artery       &\textcolor{blue}{Precision}(\%)   &\textcolor{blue}{Recall}(\%)  & \textcolor{blue}{F-measure}(\%)  & \textcolor{red}{Dice}(\%)   \\
				\hline
				 AO            &100            &100          & 100    &95.67  \\
                 CA            &100            &86.67         &92.86      &67.79  \\
                 SMA            &100            &100         &100     &86.09  \\
                 SA            &100            &100         &100     &94.42  \\
                 CHA            &100            &100         &100     &89.92  \\
                 LGA            &100            &96.67         &98.31    &89.37  \\
                 GDA            &96.55            &96.55         &96.55  &81.95   \\
                \hline
		\end{tabular}}
		\label{tablabeling}
	\end{table}

\begin{table}[htb]
        \centering
		\caption{Comparison to the existing artery labeling works, the branch-wise F-measure is shown as the evaluation metric.}
		\vspace{0pt}	
		\renewcommand\arraystretch{2}
		\setlength{\tabcolsep}{4mm}{}{
			\begin{tabular}{c|c|c|c|c|c|c|c|c}
				\hline
				Methods                  & Data    &AO    & CA    & SMA  &SA  &CHA & LGA & GDA \\
				\hline
                Oda 2012 \cite{Oda2012}   & 23    &100 &64.6   &83.3  &57.1 &33.3 &- &- \\
                Kitasaka 2017 \cite{Kitasaka2017} &50 & 95.8&93.8  &97.6 &91.1 &87.0 &89.5 &75.8        \\
                Liu 2022 \cite{Liu2022}    & 37    &98.7 &\textbf{95.5} &95.9 &84.8&88.4&-&- \\
                \hline
                \textbf{APESA (Ours)}     &30      &\textbf{100} &92.9  &\textbf{100} &\textbf{100}&\textbf{100}&\textbf{98.3}&\textbf{96.6}\\
				\hline
		\end{tabular}}
		\label{tabsotalabeling}
	\end{table}

\FloatBarrier
\section{Discussions}
We have successfully evaluated our proposed APESA method for the peripancreatic vessel segmentation and anatomical labeling. Since previous techniques have achieved nearly perfect segmentation performance on peripancreatic artery segmentation, the ITGM in our proposed APESA is designed only for the peripancreatic veins. However, this mechanism can also be applied for artery segmentation to improve the terminal error predictions. For vein segmentation, we show the experimental results by twice iterations because of the satisfactory performance on improving terminal integrity and connectivity. The experimental results show that our ITGM is an effective algorithm that helps the medical experts refine their annotations iteratively. Besides, we admit that the iterative trunk growth module may also make some mistakes such as predicting other tissues twined or adhered with the vessels in few special cases(Case@135 in Figure 12 as an example), which is the problem we aim to address in our subsequent studies.

Our proposed method follows the intuitive and effective coarse-to-fine strategy for medical segmentation. In fact, the iterative trunk growth mechanism can be trained in an end-to-end manner to find the global optimal parameters for the whole framework.  Moreover, the anatomical labeling for peripancreatic veins and other artery branches will be also further studied in our future work.

\section{Conclusions}
In this paper, we propose a novel \textbf{A}utomated \textbf{P}eripancreatic V\textbf{E}ssel \textbf{S}egmentation and L\textbf{A}beling (\textbf{APESA}) framework, for peripancreatic vein segmentation and artery labeling. Our approach is driven by two functional modules, iterative trunk growth module (ITGM) for peripancreatic vein segmentation and weakly supervised labeling mechanism (WSLM) for artery branch identification. Our proposed ITGM inspires by the prior knowledge that the vessels are fully connected and the main trunk is the most reliable prediction, it takes the largest connected component of a basic prediction as a seedling, and iteratively makes the seedling grow into a complete tree by our designed branch proposal network. Our WSLM consists of an unsupervised rule-based pseudo label generation judged by experts and an anatomical labeling network to learn the branch distribution voxel by voxel.

In our experiments, we not only improve nearly 10\% peripancreatic vein segmentation accuracy compared to the SOTA algorithms, but also boost nearly 20\% topological integrity performance compared to the previous techniques. Besides, we also achieve the best and competitive performance for peripancreatic artery segmentation and labeling, respectively.

\section*{Acknowledgement}
This work is supported by Ministry of Science and Technology of the People's Republic of China(No. 2020YFA0713800), National Natural Science Foundation of China(No. 12090023) and Postgraduate Research \& Practice Innovation Program of Jiangsu Province(No. KYCX22\_0082).

\end{document}